\documentclass[journal]{vgtc}                     
\PassOptionsToPackage{table,dvipsnames,xcdraw, svgnames}{xcolor}

\onlineid{1567}

\vgtccategory{Research}

\vgtcpapertype{Representations \& Interaction}

\usepackage{tikz}

\newcommand{\techName}[1]{\textit{Envisage}}
\newcommand{\SExpress}[1]{\textit{Query Expression}}
\newcommand{\SVerify}[1]{\textit{Query Verification}}
\newcommand{\SExecute}[1]{\textit{Query Execution}}
\newcommand{\SResult}[1]{\textit{Result Analysis}}

\definecolor{ExpressionColor}{RGB}{114, 142, 253}
\newcommand*\Expression[1]{\tikz[baseline=(char.base)]{
            \node[shape=rectangle,fill=white, draw=ExpressionColor, text=ExpressionColor, inner sep= 2pt,minimum size=11pt,rounded corners=3pt] (char) {\textbf{#1}}}}
\newcommand*\expressionA[1]{\tikz[baseline=(char.base)]{
            \node[shape=rectangle,fill=ExpressionColor, text=white, inner sep= 1.5pt,minimum size=8pt,rounded corners=0pt] (char) {\textbf{#1}}}}
\newcommand*\expressionNum[1]{\tikz[baseline=(char.base)]{
            \node[shape=circle,fill=ExpressionColor, text=white, inner sep= 0.5pt,minimum size=7pt] (char) {\textbf{#1}}}}

\definecolor{VerificationColor}{RGB}{128, 112, 181}
\newcommand*\Verification[1]{\tikz[baseline=(char.base)]{
            \node[shape=rectangle,fill=white, draw=VerificationColor, text=VerificationColor, inner sep= 2pt,minimum size=11pt,rounded corners=3pt] (char) {\textbf{#1}}}}
\newcommand*\verificationB[1]{\tikz[baseline=(char.base)]{
            \node[shape=rectangle,fill=VerificationColor, text=white, inner sep= 1.5pt,minimum size=8pt,rounded corners=0pt] (char) {\textbf{#1}}}}
\newcommand*\verificationNum[1]{\tikz[baseline=(char.base)]{
            \node[shape=circle,fill=VerificationColor, text=white, inner sep= 0.5pt,minimum size=7pt] (char) {\textbf{#1}}}}

\definecolor{ExecutionColor}{RGB}{191, 128, 146}
\newcommand*\Execution[1]{\tikz[baseline=(char.base)]{
            \node[shape=rectangle,fill=white, draw=ExecutionColor, text=ExecutionColor, inner sep= 2pt,minimum size=11pt,rounded corners=3pt] (char) {\textbf{#1}}}}
\newcommand*\executionC[1]{\tikz[baseline=(char.base)]{
            \node[shape=rectangle,fill=ExecutionColor, text=white, inner sep= 1.5pt,minimum size=8pt,rounded corners=0pt] (char) {\textbf{#1}}}}
\newcommand*\executionNum[1]{\tikz[baseline=(char.base)]{
            \node[shape=circle,fill=ExecutionColor, text=white, inner sep= 0.5pt,minimum size=7pt] (char) {\textbf{#1}}}}

\definecolor{ResultColor}{RGB}{128, 164, 146}
\newcommand*\resultD[1]{\tikz[baseline=(char.base)]{
            \node[shape=rectangle,fill=ResultColor, text=white, inner sep= 1.5pt,minimum size=8pt,rounded corners=0pt] (char) {\textbf{#1}}}}
\newcommand*\resultNum[1]{\tikz[baseline=(char.base)]{
            \node[shape=circle,fill=ResultColor, text=white, inner sep= 0.5pt,minimum size=7pt] (char) {\textbf{#1}}}}

\definecolor{CaptionColor}{RGB}{89, 91, 97}
\newcommand*\captionID[1]{\tikz[baseline=(char.base)]{
            \node[shape=rectangle,fill=CaptionColor, text=white, inner sep= 1pt,minimum size=8pt,rounded corners=1pt] (char) {\textbf{#1}}}}

\newcommand{\modify}[1]{\textcolor{black}{#1}}

\title{\techName{}: Towards Expressive Visual Graph Querying}

\author{%
  \authororcid{Xiaolin Wen}{0000-0002-8562-7640},
  \authororcid{Qishuang Fu}{0009-0000-6964-6641}, 
  \authororcid{Shuangyue Han}{0009-0001-5198-1860},
  \authororcid{Yichen Guo}{0009-0000-0721-4763},
  \authororcid{Joseph K. Liu}{0000-0001-6656-6240},
  and \authororcid{Yong Wang}{0000-0002-0092-0793}
}

\authorfooter{
  \item
  	Xiaolin Wen, Shuangyue Han, Yichen Guo, and Yong Wang are with Nanyang Technological University.
  	E-mail: \{xiaolin004, shuangyu001, yguo017\}@e.ntu.edu.sg, yong-wang@ntu.edu.sg.
  \item
  	Qishuang Fu and Joseph K. Liu are with Monash University.
  	E-mail: \{qishuang.fu, joseph.liu\}@monash.edu.
  \item Yong Wang is the co-corresponding author.
}

\abstract{%
    Graph querying is the process of retrieving information from graph data using specialized languages (e.g., Cypher), often requiring programming expertise. Visual Graph Querying (VGQ) streamlines this process by enabling users to construct and execute queries via an interactive interface without resorting to complex coding.
    \modify{However, current VGQ tools only allow users to construct simple and specific query graphs, limiting users' ability to interactively express their query intent, especially for underspecified query intent.
    To address these limitations, we propose \techName{}, an interactive visual graph querying system to enhance the expressiveness of VGQ in complex query scenarios by supporting intuitive graph structure construction and flexible parameterized rule specification.
    Specifically, \techName{} comprises four stages:} \textit{Query Expression} allows users to interactively construct graph queries through intuitive operations; \textit{Query Verification} enables the validation of constructed queries via rule verification and query instantiation;
    \textit{Progressive Query Execution} 
    can
    progressively execute queries to ensure meaningful querying results; and \textit{Result Analysis} facilitates result exploration and interpretation.
    To evaluate \techName{}, we conducted two case studies and in-depth user interviews with 14 graph analysts. The results demonstrate its effectiveness and usability in constructing, verifying, and executing complex graph queries.

}


\teaser{
  \centering
  \includegraphics[width=0.9\linewidth, alt={.}]{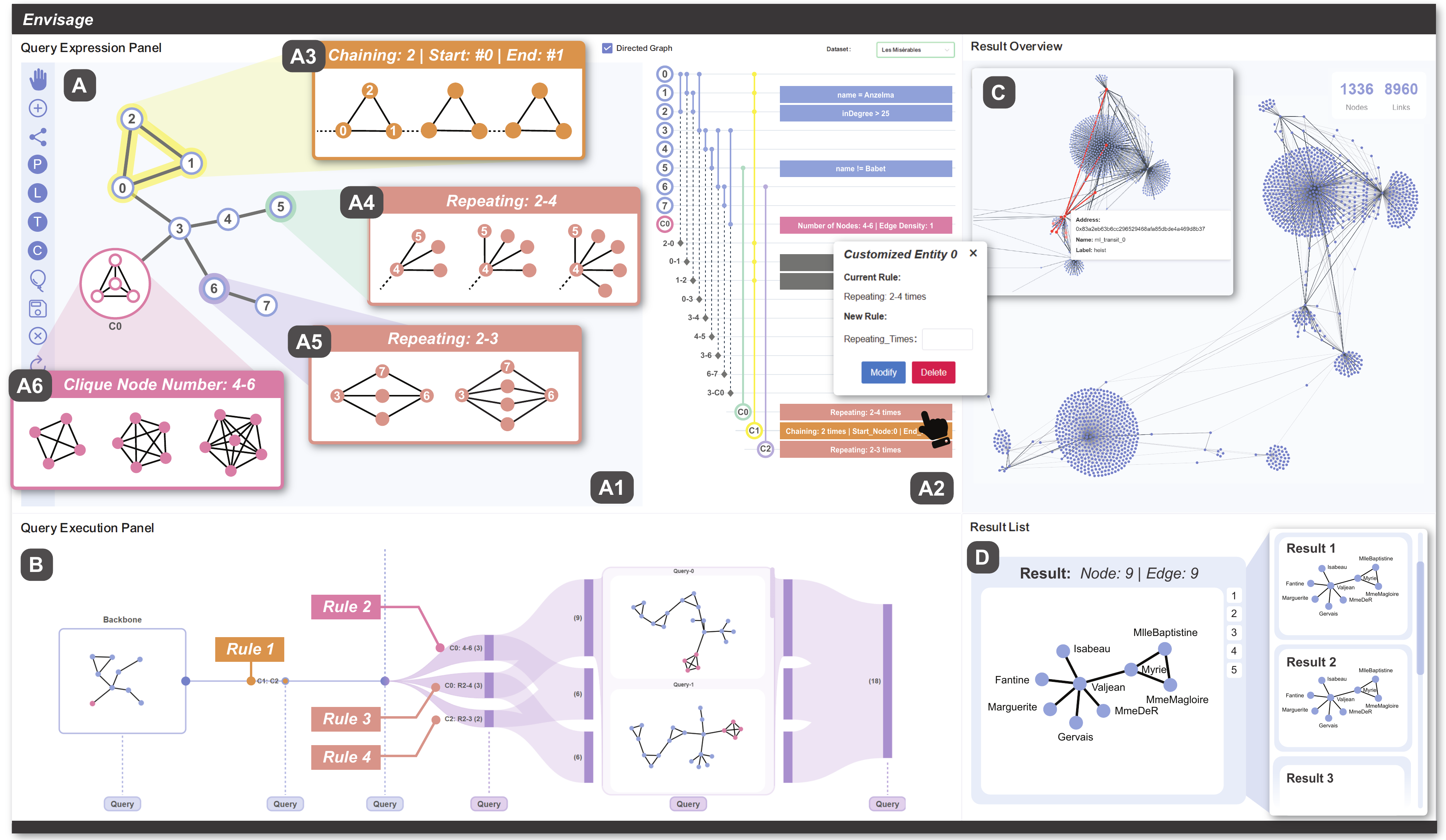}
  \caption{%
  	The Interface of \techName{} includes four components: \textit{Query Expression Panel} (A), \textit{Query Execution Panel} (B), \textit{Result Overview} (C), and \textit{Result List} (D). \modify{
Users can flexibly perform visual graph querying by quickly constructing a graph structure (A1), applying various rules (A2), verifying and executing the generated query instances (B), and analyzing the results (C and D).}
  }
  \label{fig:teaser}
}

\graphicspath{{figs/}{figures/}{pictures/}{images/}{./}} 

\usepackage{tabu}                      
\usepackage{booktabs}                  
\usepackage{lipsum}                    
\usepackage{mwe}                       

\usepackage{mathptmx}                  
\usepackage{amsmath}
\begin{document}

\maketitle
\section{Introduction} \label{sec:intro}
Graph data represents the relationships among entities and has become widely used in various application domains, including social network analysis~\cite{freeman2004development}, bioinformatics~\cite{troidl2023vimo}, and financial fraud detection~\cite{wen2023nftdisk}.
Graph querying refers to the process of retrieving relevant information (e.g., nodes, edges, and subgraphs) from graph data according to user-defined rules~\cite{yang2014schemaless}.
It is a fundamental operation for many graph exploration
tasks, such as node labeling~\cite{li2022motif}, anomaly detection~\cite{lin2024denseflow}, and pattern analysis~\cite{tamersoy2014large}.
Traditional graph querying methods typically require users to construct text-based queries formulated in specialized graph query languages such as Cypher~\cite{francis2018cypher} and GraphQL~\cite{he2008graphql}, which are subsequently executed on graph databases like Neo4j~\cite{miller2013graph}.
However, composing such queries using graph query languages typically requires strong programming skills~\cite{bhowmick2017graph}.
To address this problem, several studies have proposed leveraging visualization and human-computer interaction (HCI) techniques to simplify the query construction process~\cite{chau2008graphite,bhowmick2013vogue,pienta2017vigor,cuenca2021vertigo,troidl2023vimo}, which is referred to as \textit{\textbf{Visual Graph Querying (VGQ)}}.
Specifically, VGQ enables users to interactively construct graph queries by dragging and linking nodes~\cite{chau2008graphite,bhowmick2013vogue,pienta2016visage}, selecting predefined subgraph examples~\cite{song2021interactive}, and imposing specific constraints on graph nodes and edges~\cite{troidl2023vimo}.
\modify{Nevertheless, existing VGQ approaches primarily support simple and fully-specified graph queries, such as small, fixed subgraphs with attribute constraints.
This \textbf{limited expressiveness} restricts users' ability to construct and refine queries flexibly using such VGQ systems, particularly when addressing underspecified query intent, which is essential in many real-world scenarios~\cite{bhowmick2017graph}.
In this paper, underspecified query intent refers to a loosely defined graph pattern that can match multiple possible queries, similar to how a regular expression can match a variety of string patterns~\cite{thompson1968programming}.
For example, a user may want to query a
``connector'' subgraph (i.e., two nodes connected to a common set of intermediary nodes~\cite{dunne2013motif}, as shown in Fig.~\ref{fig:teaser}\captionID{A5}), without specifying the exact number of intermediary nodes.
}



\modify{
We surveyed existing research on VGQ~\cite{pienta2016visage,bhowmick2017graph,song2021interactive,bhowmick2022data,ma2024sierra} and conducted a preliminary study (Sec.~\ref{sec:preliminary}) with four graph analysis experts to identify the challenges in expressing graph query intent via existing VGQ systems.}
Specifically, there are four major challenges (\textbf{C1-C4}) in achieving expressive visual graph querying.
\modify{\textbf{C1. Full Expression of Underspecified Graph Query Intent.} 
Most existing VGQ tools require users to define a concrete query graph structure.
However, user intent is often underspecified and cannot be fully captured by a fixed graph structure. 
For example, a user may want to query a subgraph with two loops connected by a path, where the number of nodes in each component can be different. Such query intentions are common and can be expressed using operators like the \textit{Kleene star} in some graph query languages~\cite{egi2018loop}, but are difficult to express using existing VGQ tools. 
\textbf{C2. Fast Specification of Repetitive Graph Structure.}
Users' query intent may involve repetitive substructures, but existing VGQ tools require users to manually construct each substructure, which is inefficient and tedious.
For the aforementioned ``connector'' pattern,
it would be highly time-consuming to manually create a large number of intermediary nodes and connect them accordingly.
This becomes even more challenging when the repeated substructures are complex.
} 
\textbf{C3. Flexible Configuration of Node/Edge Attribute Constraints in Graph Queries.}
Specifying attribute constraints (e.g., node labels or edge attribute values) is essential for constructing graph queries~\cite{pienta2016visage}.
However, existing tools like SIERRA~\cite{ma2024sierra} and Visage~\cite{pienta2016visage}
typically require users to edit these constraints for each node or edge individually, which is laborious and time-consuming.
As the number of necessary constraints increases,
it becomes increasingly difficult to review, verify, and modify graph queries.
\modify{
\textbf{C4. Effective Verification and Execution of Query Instances.}
Even when users can express the aforementioned underspecified query intent, the possible number of concrete query instances reflecting their intended patterns can be enormous, which demands an effective way to verify, select, and execute appropriate graph query instances aligned with their query intent.}


To address these challenges, we
propose a novel expressive visual graph querying framework and further develop
\underline{\textbf{\techName{}}}, an interactive visual analytics system to \underline{\textbf{en}}hance the expressiveness of \underline{\textbf{vis}}u\underline{\textbf{a}}l \underline{\textbf{g}}raph qu\underline{\textbf{e}}rying.
\modify{\techName{} enables users to conduct visual graph querying through four stages: \textit{Query Expression}, \textit{Query Verification}, \textit{Progressive Query Execution}, and \textit{Result Analysis}.
Users begin by expressing their query intent, including underspecified intent (\textbf{C1}), through intuitive operations such as customized motif configuration and parameterized rule specification, where two types of rules (i.e., repeating and chaining) help users efficiently define repetitive graph patterns (\textbf{C2}).
\techName{} further enables quick configurations of attribute constraints for different entities by specifying corresponding attribute rules (\textbf{C3}).}
Then, users can check and confirm that their constructed graph query matches their intended query by verifying their defined rules and the generated query instances (\textbf{C4}).
\modify{Users can also progressively execute query instances to identify and rectify issues in the current graph query (\textbf{C4}).}
The query results are displayed as a list of subgraphs (Fig.~\ref{fig:teaser}\captionID{D}), with their distribution revealed in the input graph (Fig.~\ref{fig:teaser}\captionID{C}).
\modify{
To evaluate \techName{}, we conducted two case studies and in-depth user interviews with 14 graph analysts. The results demonstrate that \techName{} enables expressive visual graph querying, allowing users to effectively explore graph data.}
Our main contributions are as follows:

\begin{itemize}
    \item We formulate design requirements for \modify{flexibly expressing query intent through visual graph querying, based on a preliminary study with four graph analysis experts, and propose an interactive visual graph querying framework to meet the requirements.}
    
    \item We introduce \techName{}, an interactive visual graph querying system based on our proposed framework to help users express, verify, and execute graph queries in a flexible and expressive way.
    
    \item We present two case studies and conduct in-depth user interviews with 14 graph analysts to demonstrate the effectiveness and usefulness of \techName{}. 
\end{itemize}
\section{Related Work}
This work is related to prior research on \textit{visual graph querying}
and \textit{subgraph matching}.

\subsection{Visual Graph Querying}
\modify{Given our focus on effectively expressing query intent}, 
we comprehensively review the query construction processes of existing VGQ methods.
First, most VGQ systems, such as Graphite~\cite{chau2008graphite}, Prague~\cite{jin2012prague}, VOGUE~\cite{bhowmick2013vogue}, ViSual~\cite{bhowmick2015visual}, and VIMO~\cite{troidl2023vimo}, require users to manually build query graphs by interactively specifying each node and edge~\cite{bhowmick2017graph}.
One exception is VIGOR~\cite{pienta2017vigor}, which accepts text-based queries and focuses on visualizing both queries and results.
\modify{Second, some systems adopt a ``\textit{query-by-template}'' approach~\cite{pienta2016visage}, enabling users to construct queries from predefined graph templates quickly.}
For instance, VisualNeo~\cite{huang2023visualneo} enables users to construct a subgraph query with multiple nodes and edges by performing a single click-and-drag action.
Likewise, Song et al.~\cite{song2021interactive} enable users to conduct example-based graph querying by allowing them to select a graph instance to construct the query graph.
Third, specifying constraints on attributes of nodes and edges is essential for meaningful graph querying~\cite{chau2008graphite,pienta2016visage,song2021interactive}.
\modify{SIERRA~\cite{ma2024sierra} introduces visual abstraction, called labeled composite graph, to represent attribute constraints in a visual graph query. 
Also, some tools support interactions tailored to specific domain tasks and graph types, such as knowledge graphs~\cite{vargas2020user}, multilayer networks~\cite{cuenca2021vertigo}, bipartite networks~\cite{pister2022visual}, and hierarchical graphs~\cite{li2024hiregex}. 
%
%
}

\modify{
Existing methods require users to construct concrete and fully-specified query graphs, which is time-consuming and has no support for underspecified query intent that are common in real-world applications. 
\techName{} allows users to configure customized motifs and specify parameterized rules to flexibly express their queries.
Users can quickly construct repetitive graph patterns and specify attribute constraints by combining multiple rules.
Furthermore, \techName{} enables users to verify query instances derived from underspecified query intent, ensuring that query execution aligns with their expectations.
}

\subsection{Subgraph Matching}
Subgraph matching aims to find all subgraphs \( g \) in a data graph \( G \) that match a given query graph pattern \( q \). 
Existing subgraph matching algorithms can be categorized into three groups: backtracking-based, join-based, and neural network-based approaches. 
\textbf{Backtracking-based approaches} optimize depth-first search through filtering and pruning techniques.
Notable methods include VF2~\cite{cordella2004vf2}, VF2++~\cite{juettner2018vf2plusplus}, and DP-iso \cite{han2019daf}, which reduces the search space via preprocessing. 
Many graph databases integrate these techniques~\cite{shang2008taming, he2008graphql,rivero2017efficient, ren2015exploiting, han2013turboiso, zhang2009gaddi}, with GraphQL~\cite{he2008graphql} improving efficiency through local pruning and global refinement. 
\modify{\textbf{Join-based approaches} decompose queries into smaller sub-patterns and then combine intermediate results through join operations, such as binary joins or generalized multi-way joins~\cite{aberger2016emptyheaded,ammar2018bigjoin,lai2019distributed,mhedhbi2019optimizing,sun2021rapidmatch}.} 
Cypher~\cite{he2008graphql}, Neo4j’s core query language, applies this approach to property graphs.
\textbf{Neural network-based Approaches} utilize graph neural networks (GNNs) for approximate matching. GraphSAGE ~\cite{hamilton2017graphsage} improves scalability, while Neuralign~\cite{song2021interactive} builds upon it and enhances dynamic graph alignment. GNN-PE~\cite{ye2024gnnpe} introduces exact matching with path-dominance embeddings but is computationally expensive.

\modify{
User-defined queries in \techName{} are ultimately represented as individual subgraphs with attribute constraints, which can be readily translated into query languages that support fundamental features like subgraph matching, attribute filtering, and variable-length path matching. 
\techName{} adopts Cypher via Neo4j~\cite{neo4j} for its user-friendly syntax and robust support for property graph querying.}

\section{Background}
This section introduces background information, including \textit{graph querying}, \textit{graph definitions}, and \textit{motif definitions}.

\subsection{Graph Querying}\label{sec:querying}
\textit{\textbf{Graph querying}} refers to the entire process of interacting with a graph database, from formulating high-level queries to retrieving results. 
Typically, users define the desired patterns using a specialized graph query language. 
While graph databases support other query types, such as traversal queries (e.g., ``find the shortest path between two nodes''), our work focuses specifically on pattern-matching queries. 
Depending on the database (e.g., Neo4j~\cite{neo4j}, Amazon Neptune~\cite{amazon_neptune}, ArangoDB~\cite{arangodb}), users may use different query languages such as Cypher~\cite{francis2018cypher}, Gremlin~\cite{rodriguez2015gremlin}, or SPARQL~\cite{perez2009sparql} to express these patterns, which are collectively referred to as \textit{\textbf{graph queries}}. The graph database then parses and executes the graph queries, returning results in formats such as JSON, tables, or graph visualizations.
\textbf{\textit{Visual graph querying}} replaces the need to write queries in a graph query language with the construction of \modify{\textit{\textbf{query graphs}}} through interactions with a graphical interface. These query graphs, composed of nodes and edges with attribute constraints, enable users to express their desired patterns. 
The query graphs are then translated into graph query language statements and executed by the underlying graph database.
\modify{In this work, users can construct a \textbf{\textit{graph query representation}} to express their underspecified query intent, which may correspond to multiple specific query graphs.
We refer to these derived, concrete query graphs as \textit{\textbf{query instances}}.}

\subsection{Graph Definitions}
\modify{
In this work, we focus on visual graph querying over directed and undirected \textit{multivariate graphs}, where both nodes and edges can have associated attributes. A multivariate graph is defined as a tuple $G = (V, E, A_V, A_E, \Sigma)$, where $V$ is a finite set of nodes, $E$ is a set of edges, and $\Sigma$ is a finite set of edge labels that allow multiple labeled edges between the same node pairs. The functions $A_V: V \rightarrow \mathcal{D}_V$ and $A_E: E \rightarrow \mathcal{D}_E$ assign attributes to nodes and edges, where the attribute values are denoted as $\mathcal{D}_V$ and $\mathcal{D}_E$ respectively. In a \emph{directed multivariate graph}, $E \subseteq V \times V \times \Sigma$, where each edge is a tuple $(u, v, \ell)$ representing a directed edge from node $u$ to node $v$ with an optional label $\ell \in \Sigma$. In an \emph{undirected multivariate graph}, $E \subseteq \{ \{u, v\} \mid u, v \in V \} \times \Sigma$, where each edge connects an unordered node pair with label $\ell \in \Sigma$. }


\subsection{Motif Definitions} \label{sec: motif}
A \textbf{\textit{motif}} refers to a small subgraph pattern that captures meaningful structural configurations within graphs. Such patterns have been extensively studied for their importance in understanding graph structures~\cite{zhou24adamotif}. In this work, we introduce a quick configuration feature for four commonly used motif types (i.e., \textit{path}, \textit{loop}, \textit{tree}, and \textit{clique}) to help users easily \modify{construct their graph queries}.
We selected these four motif types due to their frequent use in existing motif-based network visualization research~\cite{huang05motif, von2009visual,cakmak2022motif-based,jung2024monetexplorer}, aligning with our focus on visual graph querying.
\modify{
The four motif types are described below:}

\begin{itemize}
    \item \textbf{Path:} A path $P$ is defined as a sequence of vertices $v_1,v_2,\dots,v_k$ (with $k \geq 2$) such that: (1) each consecutive pair of vertices $v_i$ and $v_{i+1}$ connected by an edge $(v_i,v_{i+1}) \in E$ for $i=1,2,\dots,k-1$; (2) in undirected graphs, the edge may also be $(v_{i+1},v_i) \in E$); and (3) all vertices are distinct, i.e., $v_i \neq v_j$ for all $i \neq j$. 
    In our approach, users can specify a desired range for the number of vertices in the path to support flexible query construction.
    \item \textbf{Loop:} A loop \( L \) is a closed path in graph \( G \), defined by a sequence of vertices \( v_1, v_2, \dots, v_k \) (with \( k \geq 3 \)) such that: (1) \( v_1 = v_k \); (2) all intermediate vertices are distinct, i.e., \( v_i \neq v_j \) for all \( i \neq j \), except \( i = 1, j = k \); and (3) each consecutive pair is connected by an edge \( (v_i, v_{i+1}) \in E \) for \( i = 1, \dots, k - 1 \). In undirected graphs, edges may also be \( (v_{i+1}, v_i) \in E \). Users can specify a range for the number of vertices in the loop.
    \item \textbf{Tree:} A tree \( T \) is a connected, acyclic subgraph with a designated root node \( r \), defined over a set of vertices \( v_1, v_2, \ldots, v_k \) (with \( k \geq 2 \)) such that: (1) a unique simple path exists between any pair of vertices \( u, v \in V_T \); and (2) the number of edges satisfies \( |E_T| = |V_T| - 1 \). In directed graphs (i.e., directed trees), all edges must follow a consistent direction from parent to child (i.e., \( (v_i, v_j) \in E \) such that \( v_i \) is the parent of \( v_j \)). Users can specify a range for the number of vertices in the tree, as well as the width and depth. 
    \item \textbf{Clique:} A clique \( C \) is a complete subgraph defined over a set of vertices \( v_1, v_2, \ldots, v_k \) (with \( k \geq 2 \)) such that:  
    (1) for every pair of distinct vertices \( u, v \in V_C \), an edge \( (u, v) \in E \) exists;  
    (2) in undirected graphs, edges are symmetric, i.e., \( (u, v) \in E \) or \( (v, u) \in E \);  
    (3) in directed graphs, both edges \( (u, v) \in E \) and \( (v, u) \in E \) must exist for every pair.  
    Users can specify a desired range for the number of vertices in the clique.
\end{itemize}

\section{Informing the design}
This section presents our preliminary study and design requirements.

\subsection{Preliminary Study}~\label{sec:preliminary}
We conducted a preliminary study to comprehensively gather users' requirements regarding the expressiveness of visual graph querying. 
The details of the participants and procedures are as follows:

\textbf{Participants:} 
We invited four graph analysis experts (\textbf{E1–E4}) as participants. 
All of them are researchers from universities, each with over three years of experience in graph analysis across various domains, where retrieving subgraphs from graph data is a routine task.
\textbf{E1} and \textbf{E2} possess expertise in querying using graph databases, while \textbf{E3} and \textbf{E4} analyze graphs through programming and were unfamiliar with graph query languages.
Their feedback is also valuable because our goal is to enable users to conduct graph querying interactively without using complex query languages.

\textbf{Procedures:} The preliminary study was conducted in two sessions. 
In the first session, we conducted open-ended interviews with each participant. 
Initially, participants were asked to describe typical requirements for querying patterns within graph data in their routine research activities.
\modify{We then introduced existing graph querying methods, including both query language-based and interactive visualization approaches.
Participants were encouraged to describe any current or potential challenges in applying these methods to express and execute their previously-described requirements.}
In response to these challenges, we formulated initial design requirements and proposed an interactive framework for editing and executing graph queries.
In the second session, participants were invited to evaluate these design requirements and the proposed framework.
Their feedback guided us in finalizing the design requirements and refining the interactive framework.

\modify{
The challenges (C1-C4) presented in Section~\ref{sec:intro} are summarized from the challenges proposed by participants. The design requirements and interactive framework are detailed in Section~\ref{sec:design_requirement} and Section~\ref{sec:framework}.
}

\begin{figure*}[!htbp]
  \centering 
      \setlength{\belowcaptionskip}{-0.4cm}
  \includegraphics[width=0.9\linewidth
  ]{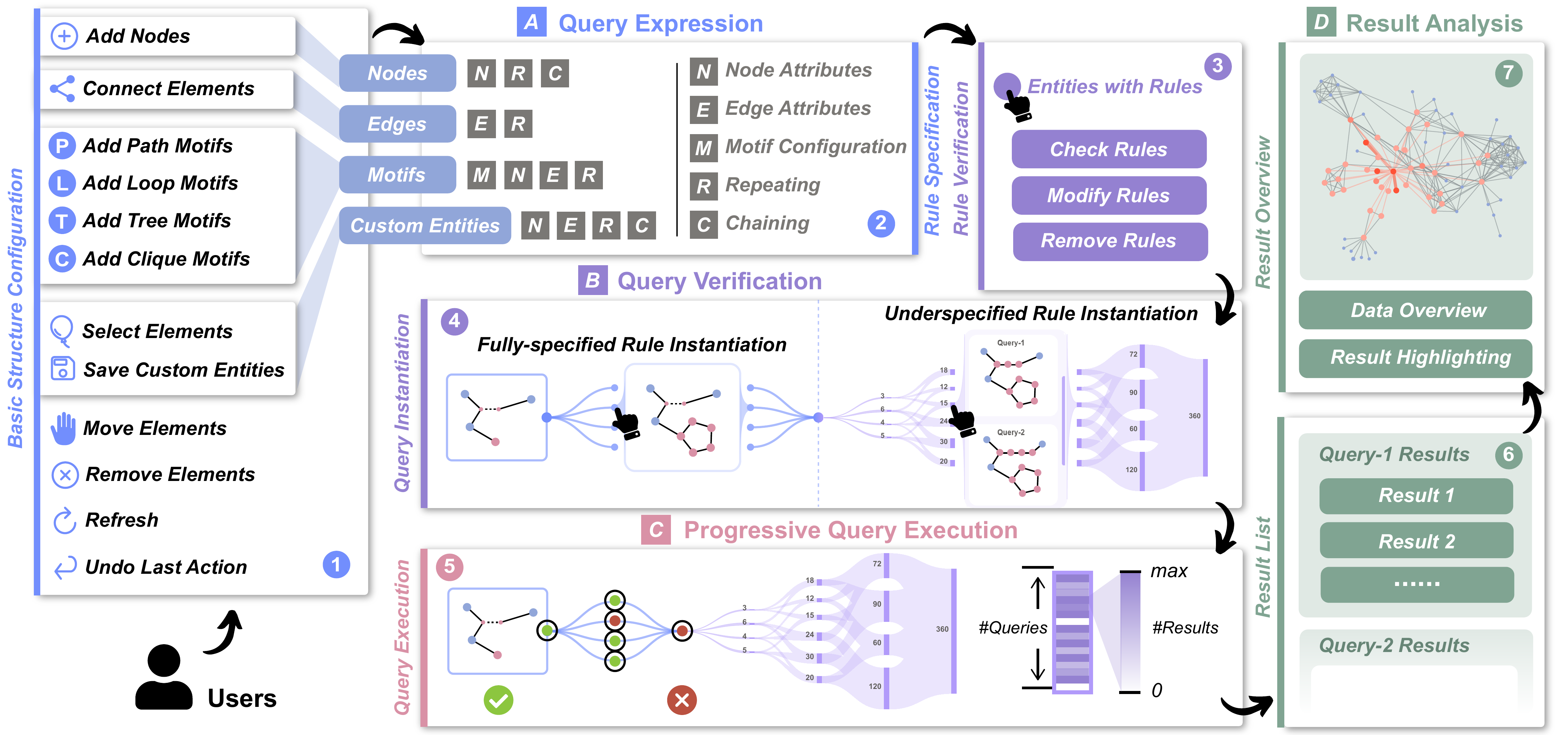}
  \caption{The \techName{} framework consists of four stages: (A) Query Expression, where users construct the basic structure of the query (1) and specify rules (2); (B) Query Verification, which allows users to inspect applied rules (3) and preview query instantiations (4); (C) Progressive Query Execution, where queries are progressively executed and refined based on intermediate results (5); and (D) Result Analysis, which presents the query outcomes through a detailed result list (6) and an overview visualization (7). }
  \label{fig:framework}
\end{figure*}

\subsection{Design Requirements}\label{sec:design_requirement}

We derived six design requirements (\textbf{R1-R6}) and grouped them into three categories: Query Expression (\Expression{Expression}), Query Verification (\Verification{Verification}), and Query Execution (\Execution{Execution}).

\begin{itemize}
    \item{\textbf{R1}} \Expression{Expression} \modify{\textbf{Enable users to specify graph structure efficiently.} All experts (\textbf{E1-E4}) noted that constructing query graphs with numerous nodes and edges is time-consuming and could be simplified. As discussed in C1, user query intent may involve motifs (e.g, loops and paths) with underspecified parameters (e.g., the number of nodes), so our approach should allow users to specify such motifs efficiently. In addition, as manually creating repetitive structures is tedious (C2), our approach should also support expressing repetitive substructures quickly.
    }

    
    \item \textbf{R2} \Expression{Expression} \modify{\textbf{Support underspecified query intent.} All experts (\textbf{E1-E4}) mentioned that a fixed query graph is insufficient to specify graph patterns that they wish to query (C1). For instance, in anomaly detection tasks, anomaly patterns often lack precise and concrete structures, and are typically
    expressed as soft structure constraints.
    The ideal query results are diverse and may vary in aspects such as node count or topology. Our approach should allow users to express such underspecified query intent, e.g., specifying motifs by approximate node counts and defining repetition ranges for particular structures.}
    
    \item  \textbf{R3} \Expression{Expression} \textbf{Facilitate the expression and review of attribute constraints.} 
    Attribute constraints are crucial in querying multivariate graphs~\cite{ma2024sierra}, and all experts (\textbf{E1–E4}) agreed on the importance of specifying attribute constraints in queries. \modify{However, existing methods require users to assign constraints individually to nodes and edges, making batch specification difficult (C3).} Our approach should allow users to easily specify constraints across multiple elements.
    Additionally, \textbf{E3} suggested providing an overview to facilitate constraint review and modification. 

    \item  \textbf{R4} \Verification{Verification} \textbf{Provide detailed query instances for verification.} \modify{Three experts (\textbf{E1-E3}) expected that an effective approach should
    allow them to
    instantiate detailed query graphs for verification (C4). \textbf{E1} emphasized that replacing specific query graphs with underspecified query expressions might result in queries that do not align with user expectations. Therefore, our approach should instantiate user-defined expressions into concrete graphs for an easy verification by users.}
    
    \item  \textbf{R5} \Verification{Verification} \textbf{Offer guidance for query modification.} 
    A user-friendly visual query system should guide users in constructing correct queries~\cite{bhowmick2017graph}.
    \modify{Unexpected query instances or empty query results may be caused by issues in certain parts of a query expression.
    \textbf{E1} and \textbf{E2} suggested that our approach should offer a clear guidance to identify which parts deviate from their expectations, facilitating easy modifications.}
    
    \item  \textbf{R6} \Execution{Execution} \textbf{Execute user-defined queries and display the results.} \modify{All experts (\textbf{E1-E4}) emphasized that our approach should promptly execute them and display the results (C4). \textbf{E1} and \textbf{E2} suggested transforming user-defined query expressions into graph query language formats and executing them through a graph database to support precise subgraph matching.} Additionally, \textbf{E4} noted that presenting query results within the context of the entire graph data helps users better understand result distributions and gain deeper insights for further query refinement.

\end{itemize}

\begin{figure*}[!htbp]
  \centering 
  \includegraphics[width=0.9\linewidth
  ]{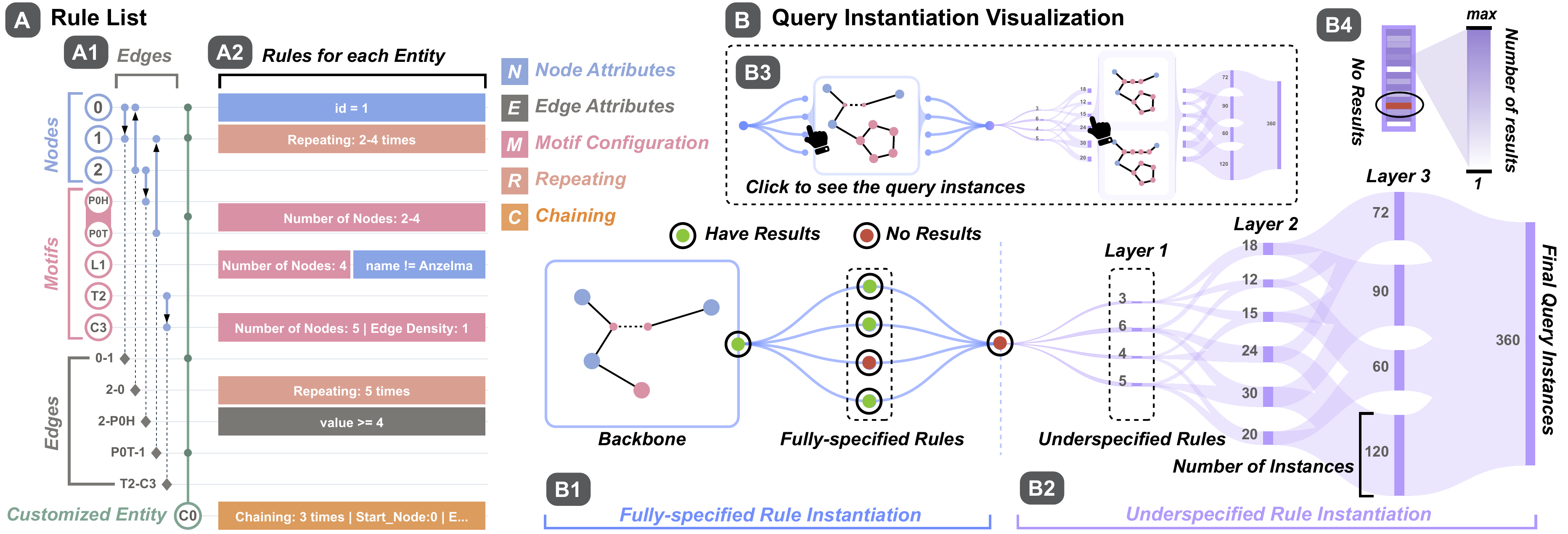}
    \setlength{\belowcaptionskip}{-0.4cm}
  \caption{The visual design of (A) the \textit{Rule List} and (B) the \textit{Query Instantiation Visualization}. The \textit{Rule List} displays all user-defined entities along the y-axis (A1), with their corresponding rules listed alongside each entity (A2). The \textit{Query Instantiation Visualization} illustrates how the graph query representation is instantiated into specific query instances based on both \modify{fully-specified} rules (B1) and \modify{underspecified} rules (B2). Users can interactively explore each query instance (B3), and the corresponding query results can be displayed within this view (B4). }
  \label{fig:design}
\end{figure*}

\section{\techName{}}\label{sec:framework}
\modify{
Informed by the above design requirements, we propose an interactive visual graph querying framework (Fig.~\ref{fig:framework}) that supports efficient expression, verification, and execution of graph queries,}
which comprises four stages: \textit{Query Expression}, \textit{Query Verification}, \textit{Progressive Query Execution}, and \textit{Result Analysis}. 
\modify{We implemented this framework in a prototype system, \techName{}\footnote{Video demo available at \url{https://youtu.be/d1KR5McgWPA}} (Fig.~\ref{fig:teaser}), using JavaScript (Vue.js and D3.js) and backed by a Neo4j graph database.
This section outlines each stage and the corresponding user interactions in \techName{}.}

\subsection{Query Expression} \label{sec: expression}
\modify{
The \SExpress{} stage (Fig.~\ref{fig:framework}\expressionA{A}) enables users to express their query intent through two phases: \textit{basic structure configuration} and \textit{rule specification}.
In addition to standard graph construction, it supports efficient expression of underspecified query intent via advanced functions like customized motif configuration and parameterized rules. 
}


\textbf{Basic Structure Configuration:}
Initially, users can configure the basic structure of a 
query graph
through a set of interactions (as shown in Fig.~\ref{fig:framework}\expressionNum{1}), including four types of entities: nodes, edges, motifs, and customized entities.
Motifs represent four common graph structures described in Section~\ref{sec: motif}, allowing users to quickly build queries with specific patterns. 
For clarity and consistency, all external edges connected to motifs, except paths, are treated as connecting to a representative node within the motif (e.g., the root of a tree). Paths, however, expose both a head and a tail node for connection.
A customized entity comprises a group of nodes, edges, and motifs marked by users, enabling batch operations such as setting attribute constraints.


\textbf{Rule Specification:}
To support efficient query expression, we introduce five types of \modify{parameterized} rules that can be applied to entities defined during the basic structure configuration stage. 
These rules enable users to specify structural or semantic constraints on nodes, edges, motifs, and customized entities within the query graph. As illustrated in Fig.~\ref{fig:framework}\expressionNum{2}, users can interactively assign rules to relevant entities to express specific query requirements. The definitions and functionalities of the five rule types are described in detail below.

\begin{itemize}
    \item \textbf{Node Attribute Rules} specify attribute constraints for nodes within an entity, such as requiring a numerical attribute to exceed a threshold. These rules apply to individual nodes as well as to nodes within motifs or customized entities.
    \item \textbf{Edge Attribute Rules} specify attribute constraints for edges within an entity, such as requiring edge weights to fall within a certain range. These rules apply to individual edges as well as to edges within motifs or customized entities.
    \item \textbf{Motif Configuration Rules} \modify{define the structure of motifs using parameters (e.g., number of nodes). 
    To express the underspecified intent, users can assign value ranges to these parameters.} Fig.~\ref{fig:teaser}\captionID{A6} illustrates several subgraph instances that conform to a clique motif configuration with a specified node range of 4 to 6.
    
    \item \textbf{Repeating Rules} define how a substructure is repeated within the query graph \modify{by specifying parameters such as the number of repetitions and the entities to repeat. These rules are applicable to all entity types.}
    When applied to a customized entity, the rule duplicates all included nodes, edges, and motifs and reconnects duplicated components to the same external entities as the original, thereby preserving the structural context.
    Similarly, repeating rules for nodes and motifs duplicate them while preserving external connections, whereas edge repetition duplicates only the edges themselves without duplicating the connected nodes. Fig.~\ref{fig:teaser}\captionID{A4} and \captionID{A5} showcase the use of repeating rules on a single node to express a star structure and a connector pattern, respectively.
    
    \item \textbf{Chaining Rules} \modify{define how a substructure is duplicated and connected into a chain-like pattern, which applies to both nodes and customized entities. Users can configure chaining rules by selecting a customized entity or node and specifying parameters, such as the start node, end node, number of chaining iterations, and chaining mode.
    Chaining rules duplicate the selected entities and connect the duplicated entities sequentially.} 
    The \textit{chaining mode} determines how each duplication is connected: either by linking the start node of the new duplication to the end node of the previous duplication (Fig.~\ref{fig:teaser}\captionID{A3}), or by using the end node of each duplication as the start node for the next iteration (Fig.~\ref{fig:case_ml}\captionID{B1}).
\end{itemize}

\textit{Node Attribute Rules} and \textit{Edge Attribute Rules} are fundamental features in some existing VGQ tools~\cite{pienta2016visage,troidl2023vimo}. However, our approach allows users to batch-apply these rules to customized entities and motifs (\textbf{R3}). 
\modify{\textit{Motif Configuration Rules} enable users to efficiently construct query graphs with specific structures (\textbf{R1}) and express the underspecified intent regarding the size and shape of motifs (\textbf{R2}). \textit{Repeating Rules} and \textit{Chaining Rules} facilitate the rapid specification of the query graphs containing repetitive patterns (\textbf{R1}), and also allow users to express underspecified intent by specifying a range for the parameters (\textbf{R2}).
To distinguish the users' query expressions (i.e., basic structures with parameterized rules) from concrete query instances, we refer to them as \textit{\textbf{graph query representations}}, as mentioned in Section~\ref{sec:querying}.}

\textbf{Interactions with \techName{}:}
\techName{} provides a \textit{Query Expression Panel} (Fig.~\ref{fig:teaser}\captionID{A}) that allows users to construct their graph query representations. The panel consists of two main components: the \textit{Query Editor} (Fig.~\ref{fig:teaser}\captionID{A1}) and the \textit{Rule List} (Fig.~\ref{fig:teaser}\captionID{A2}).
\modify{
In the \textit{Query Editor}, users can add nodes and motifs by clicking buttons and reposition them via drag-and-drop. By clicking the ``\textit{Connect Elements}'' button, users can draw edges between two entities.}
To create customized entities, users can select multiple entities by clicking on them and then finalize the selection by clicking the ``\textit{Save}'' button. 
\modify{For rule specification, users can right-click on individual entities to open a pop-up menu, where rule parameters can be configured.
The \textit{Rule List} displays all entities and the corresponding rules applied to them, which will be described in detail in Section~\ref{sec:verification}.
For entities that are difficult to select directly in the \textit{Query Editor} (e.g., customized entities), users can instead right-click on their labels in the \textit{Rule List} to access the same pop-up menu. 
}

\subsection{Query Verification}\label{sec:verification}
\textit{Query Verification} (Fig.~\ref{fig:framework}\verificationB{B}) is designed to help users confirm whether their query graph representations align with their expectations, which consists of \textit{Rule Verification} and \textit{Query Instantiation Verification}.

\textbf{Rule Verification:} 
\modify{
Before executing queries, users can check whether the applied rules accurately reflect their query intent, and then modify or remove any incorrect or unintended rules as needed (Fig.~\ref{fig:framework}\verificationNum{3}), using the \textit{Rule List} (Fig.~\ref{fig:design}\captionID{A}).
To facilitate rule verification, all entities are listed along the y-axis and grouped by type, while the rules applied to each entity are displayed as colored blocks with text, arranged horizontally to the right (Fig.~\ref{fig:design}\captionID{A2}).
Different colors indicate different rule types, making it easier for users to distinguish and review them.
By right-clicking a colored block, users can open a pop-up menu to modify and remove the corresponding rules.
To help users identify which entities are connected by edges or included in customized entities, we offset the label positions of edges and customized entities, draw vertical lines linking them to their corresponding entities, and add arrows to edges to indicate the direction from source to target.
This design is inspired by \textit{Massive Sequence View}~\cite{van2013dynamic} and \textit{BioFabric}~\cite{fuchs2024exploring}, which effectively encodes graph structural information in node-list layouts.
Additionally, users can hover over an entity to highlight it and its related entities in both \textit{Query Editor} and \textit{Rule List}, establishing visual connections between the two components.
}


\textbf{Query Instantiation Verification: }
In line with \textbf{R4}, \modify{we generate all valid query instances based on user-defined query representations for verification.
To help users understand how these instances are derived and identify any misalignment with their query intent (\textbf{R5}), we visually present the \textit{query instantiation} process (Fig.~\ref{fig:framework}\verificationNum{4}) based on user-defined rules, which includes two phases: \textit{\textbf{fully-specified rule instantiation}} and \textit{\textbf{underspecified rule instantiation}}.
\textit{Fully-specified rules} are those in which all parameters are fixed values, resulting in a single query instance.
In contrast, \textit{underspecified rules} include at least one parameter defined as a value range, capturing flexible user intent and producing multiple possible query instances.
The above definitions only relate to three of the four rule types (i.e., motif configuration, repeating, and chaining). Attribute rules do not affect instance variation and are directly assigned to the relevant nodes and edges in query instances.} The details of both phases are discussed below:


\begin{itemize}
     \item \modify{\textbf{Fully-specified Rule Instantiation:} 
     We first construct a \textit{backbone} based on users' query representations, where entities with repeating and chaining rules appear only once.
    Motifs other than paths are abstracted as single nodes, while paths are represented by two nodes (a head node and a tail node) connected by a path. This abstraction preserves the fundamental structure of user-defined queries, making them easier to understand as a starting point for instantiation.
    We begin by expanding the backbone using each fully-specified rule, then apply all these rules collectively, and finally visualize the resulting query instances for users. This visualization helps users understand the final fully-specified query instance while allowing them to inspect individual rules to guide modifications (\textbf{R5}).}
    
    \item \modify{\textbf{Underspecified Rule Instantiation:} 
    Each underspecified rule can generate multiple possible query instances, and combining several such rules can lead to a large number of combinations.
    To manage this complexity, we start from the final fully-specified query instance and incrementally generate all possible combinations of underspecified rules and their corresponding instances.
    This process follows a layered strategy: each layer represents instances generated by applying a specific number of underspecified rules. 
    For example, Layer 2 includes all instances formed by applying any two underspecified rules to the fully specified base. The final layer includes instances that satisfy all user-defined rules.
    This layer-by-layer organization allows users to efficiently review the query generation process (\textbf{R4}) and identify rule combinations that cause unintended outcomes for further modification (\textbf{R5}).}

\end{itemize}

During the \textit{query instantiation} process, motifs are expanded into specific nodes and edges, while entities with repeating or chaining rules are duplicated according to the rule descriptions in Section~\ref{sec: expression}.
The query instances consist solely of nodes and edges, with their \textit{node attribute rules} and \textit{edge attribute rules} accordingly.

\textbf{Interactions with \techName{}:} 
\modify{
For \textit{rule verification}, users can begin by browsing all applied rules (colored blocks with text) associated with each entity. By hovering over an entity of interest, users can highlight the corresponding rules and related entities in both the \textit{Rule List} and the \textit{Query Editor}, enhancing the clarity. When finding unintended rules, users can right-click on the colored blocks and edit or remove them in the pop-up menu (Fig.~\ref{fig:teaser}\captionID{A2}).
For \textit{query instantiation verification}, \techName{} provides a \textit{\textbf{query instantiation visualization}} (Fig.~\ref{fig:design}\captionID{B}) to illustrate the instantiation process.
On the left, a node-link diagram shows the backbone structure (Fig.~\ref{fig:design}\captionID{B1}) linked to several circles, with each circle representing a fully-specified rule.
Clicking a circle shows the corresponding query instance in another node-link diagram (Fig.~\ref{fig:design}\captionID{B3}), allowing users to verify whether each rule expands the backbone as intended. 
These circles converge into a single circle representing the final fully-specified query instance, which can also be clicked to review and confirm whether all fully-specified rules are correctly applied.
This final instance is further linked to a Sankey diagram-based design to display the \textit{underspecified rule instantiation} (Fig.~\ref{fig:design}\captionID{B2}). 
The x-axis layer index indicates the number of underspecified rules selected in each combination. For example, Layer 1 includes individual rules, Layer 3 includes all combinations of three rules from the full rule set, and the final layer represents the full combination of all rules.
Each rectangle within a layer represents a rule combination, with its height indicating the number of query instances generated by that combination.
The flows between adjacent layers indicate that each combination in the next layer includes multiple combinations from the previous layer. For example, the combination [0,1,2] in Layer 3 contains the subsets [0,1], [0,2], and [1,2].
By clicking these rectangles, users can review and compare the query instances generated by different rule combinations to identify which underspecified rules produce unintended results (Fig.~\ref{fig:design}\captionID{B3}).
Additionally, rule descriptions and the number of instances are displayed in text alongside corresponding visual elements for clarity.}

\subsection{Progressive Query Execution}\label{sec: execution}
\modify{The generated query instances exist as graph structures with attribute constraints, rather than in graph query language statements that are directly executable by the graph database.
To support execution, we introduce a \textit{query translator} that converts these instances into graph query language statements.}
However, executing all final query instances at once can be time-consuming. 
\modify{If the results are unavailable or unexpected, users must iteratively modify the queries and re-execute them, making it difficult to diagnose issues and guide modifications (\textbf{R5}).} 
To ensure smooth execution of user-defined query instances (\textbf{R6}), we introduce a \textit{progressive execution} process (Fig.~\ref{fig:framework}\executionC{C}).
\modify{The \textit{query instantiation visualization} contains multiple intermediate steps: the backbone, individual fully-specified rules, the final specific instance, and all layers of underspecified rule combinations.
Each step builds on the results of the previous one, forming a structured and traceable execution flow.
By executing queries progressively, users can easily identify which rules cause failures. If a step returns no results, all subsequent instances depending on it are also guaranteed to fail (Fig.~\ref{fig:framework}\executionNum{5}). To speed up this process, users can initially limit the number of returned results and execute the steps incrementally, allowing them to quickly identify and adjust problematic rules. Once the queries are refined, users can execute the final queries to retrieve the complete set of results.
}


\textbf{Interactions with \techName{}:} 
\modify{The progressive query execution workflow is integrated into the \textit{Query Execution Panel} via \textit{query buttons} (Fig.~\ref{fig:teaser}\captionID{B}).}
Each intermediate step has a query button placed below it, enabling users to execute queries and view the results.
When users click the query button for an individual query instance (represented by circles), 
\techName{} translates it into query language through the query translator and executes it in Neo4j database~\cite{neo4j}.
\modify{The result is visually encoded: a green circle indicates a successful match, while a red circle indicates no results} (Fig.~\ref{fig:design}\captionID{B1}).
The number of results is displayed alongside the circle as text.
\modify{In the Sankey diagram-based design, each rectangle represents a combination of multiple underspecified rules and may include several query instances, each with its own result set.
Since the height of the rectangle reflects the number of query instances, we subdivide it into smaller vertically-arranged rectangles, each representing a single query instance. }
To enhance interpretability, we use a gradient purple color to visually encode the number of results for each query instance. \modify{The instances with no results are shown in red (Fig.~\ref{fig:design}\captionID{B4}).}
\modify{Users can selectively execute query instances at any intermediate step with a limited number of returned results, enabling them to quickly assess whether queries are functioning as intended and iteratively refine rules to ensure meaningful results.}

\begin{figure*}[!htbp]
  \centering 
  \setlength{\abovecaptionskip}{-0.05cm}
  \setlength{\belowcaptionskip}{-0.4cm}
  \includegraphics[width=\linewidth
  ]{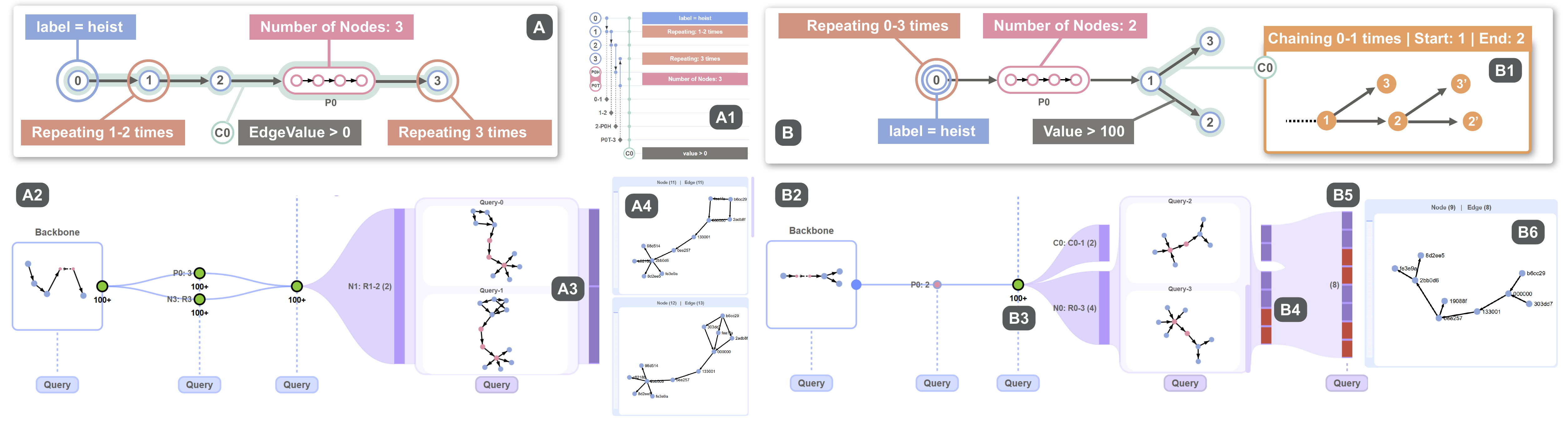}
  \caption{\modify{Money laundering pattern detection using \techName{}. (A) The user defined a dispersion–convergence–dispersion pattern by applying multiple rules (A1), executed the query progressively (A2–A3), and successfully retrieved matching results (A4). (B) She constructed a chain-shaped pattern using chaining and repeating rules (B1) and used progressive execution (B2–B4) to identify over-repetition issues as shown in the results (B5-B6).}}
  \label{fig:case_ml}
\end{figure*}

\subsection{Result Analysis}
In \textit{progressive query execution}, users can determine whether query instances produce results but still require an in-depth analysis of each instance’s results (\textbf{R6}). The Result Analysis stage (Fig.~\ref{fig:framework}\resultD{D}) aims to present results through both an overview and a detailed examination.
In the \textit{result overview} (Fig.~\ref{fig:framework}\resultNum{7}), we first compute the frequency of all nodes and edges based on the results that users wish to analyze. These frequencies are then visualized and highlighted within the context of the entire graph data. This overview enables users to identify result distributions across the entire graph and detect potential issues, such as an excessive concentration of results within a single subgraph.
Additionally, users can examine individual query results in detail, including the graph structure and its associated properties (Fig.~\ref{fig:framework}\resultNum{6}).

\textbf{Interactions with \techName{}:} 
\modify{
\techName{} provides a \textit{Result Overview} (Fig.~\ref{fig:teaser}\captionID{C}) and a \textit{Result List} (Fig.~\ref{fig:teaser}\captionID{D}) that display the corresponding query results when users click on circles or rectangles representing query instances that have returned results.}
Specifically, the \textit{Result Overview} presents the entire graph as a node-link diagram, with result frequencies highlighted using a gradient red color scheme. 
\modify{
The \textit{Result List} displays information cards, each representing the results of a single query instance.
Each card includes statistics (e.g., node count), a node-link diagram showing one result, and a sidebar for switching between results of that instance.
Only one result is shown at a time, as all results for a given instance share the same graph structure but differ in labels. 
Users can browse these labels via the sidebar (Fig.~\ref{fig:teaser}\captionID{D}) and navigate between different instances using a scroll bar.
%
%
}


\section{Case Study} \label{sec: casestudy}
This section presents two case studies demonstrating the effectiveness of \techName{}.
To evaluate \techName{}, we conducted user interviews with 14 graph analysts (\textbf{U1-U14}), which will be discussed in Section~\ref{sec: interview}.
The cases in this section illustrate how \textbf{U1} and \textbf{U9} utilized \techName{} for specific querying tasks. 
To assess the generalizability of \techName{}, we incorporated three graph datasets of varying scales and from different domains for user selection to conduct the graph querying. The details of these datasets are described as follows:

\begin{itemize}
    \item \textbf{Les Mis{\'e}rables Co-occurrence Network (LMCN):} 
    This dataset involves co-occurrence relationships between characters in Victor Hugo's novel Les Mis{\'e}rables~\cite{WikipediaLesMis}. It is a classical undirected graph dataset for graph analysis~\cite{knuth1993stanford}, consisting of 77 nodes and 254 edges. Each node corresponds to a character, and an edge connects two characters if they appear in the same chapter.
    \item \textbf{Function Call Graph (FCG):} We select four samples from the public function call dataset,\textit{ Malicious Webshell Family (MWF)}~\cite{zhao2024malicious}, to construct this dataset, which comprises 305 nodes and 4,516 edges. Nodes represent functions, while edges denote function calls with multiple attributes, including the caller function, callee function, parameters, and return values. 
    \item  \textbf{Money Laundering Network (MLN):} This dataset captures an Ethereum-based money laundering event, named \textit{eazyfihacker}~\cite{halborn2021explained}, selected from the public Ethereum Money Laundering dataset, \textit{EthereumHeist}~\cite{wu2023toward}. It contains 1,335 nodes and 8,960 edges, where nodes represent blockchain accounts and edges denote transactions between them.

\end{itemize}

\subsection{Case 1: In-depth Analysis of Money Laundering Patterns}
\textbf{U1} expressed a particular interest in the money laundering dataset (MLN) and aimed to leverage \techName{} to identify characteristic patterns associated with money laundering activities. 

\textbf{Layering–Integration–Layering:} As a first step, \textbf{U1} sought to investigate the presence of Layering–Integration–Layering patterns within the network, which means that illicit funds are initially dispersed across multiple accounts (layering), consolidated into a single account (integration), and then dispersed again (layering). To analyze this pattern, she constructed a query representation, as illustrated in Fig.~\ref{fig:case_ml}\captionID{A}.
Specifically, she applied a node attribute rule ``\textit{label=heist}'' (\textit{heist} label marks the source node in MLN dataset)  to the initial node (\textit{Node 0}) to mark it as the suspected origin of the money laundering flow.
She then added two additional nodes (\textit{Node 1} and \textit{Node 2}), connecting them sequentially to model the initial money flow.
To express the Layering–Integration structure, she applied a repeating rule to \textit{Node 1}, allowing it to repeat 1–2 times.
To capture subsequent layering, she appended a directed path (\textit{P0}) consisting of three nodes and added \textit{Node 3} to represent the further flow of laundered funds. 
A repeating rule was then applied to \textit{Node 3} to express the final layering process.
She then grouped all entities into a customized entity (\textit{C0}) and applied an edge attribute rule, ``\textit{value>0}'', to ensure only transactions with positive monetary value were considered.
After constructing the query, \textbf{U1} reviewed the defined rules (Fig.~\ref{fig:case_ml}\captionID{A1}) and the generated query instances (Fig.~\ref{fig:case_ml}\captionID{A3}) to verify that the structure aligned with her intended pattern.
She then executed intermediate steps and the final query instances by interacting with the query interface (Fig.~\ref{fig:case_ml}\captionID{A2}). The appearance of green circles and purple blocks indicated that each step returned results successfully. Finally, in the \textit{Result List} (Fig.~\ref{fig:case_ml}\captionID{A4}), she confirmed that the dataset did indeed contain instances matching the Layering–Integration–Layering pattern of money laundering behavior.

\textbf{Chain-shaped Money Laundering: } 
\textbf{U1} then turned her attention to investigating a chain-shaped money laundering pattern and constructed a corresponding query representation, as shown in Fig.~\ref{fig:case_ml}\captionID{B}. Similar to the previous pattern, she designated \textit{Node 0} as the source node and added a directed path (\textit{P0}) to represent the primary flow of illicit funds. To simulate branching behavior within the laundering process, she added three additional nodes (\textit{Node 1}, \textit{Node 2}, and \textit{Node 3}), thereby creating two diverging branches from the main path.
She grouped these nodes into a customized entity (\textit{C0}) and applied a chaining rule, specifying \textit{Node 1} as the start node and \textit{Node 2} as the end node, to expand the query in a chain-like structure, as illustrated in Fig.~\ref{fig:case_ml}\captionID{B1}. Then, she applied an attribute rule, ``\textit{value>100}'', to the edge from \textit{Node 0} to \textit{Node 1}, marking it as the main branch to differentiate it from auxiliary paths.
Additionally, \textbf{U1} hypothesized that multiple source nodes might initiate transactions along this chain-shaped pattern, and she applied a repeating rule to the source node (\textit{Node 0}). After executing the final query instances, she observed that half of them returned results (Fig.~\ref{fig:case_ml}\captionID{B6}), while the other half returned no results, indicated by red rectangles in Fig.~\ref{fig:case_ml}\captionID{B5}.
To diagnose the issue, \textbf{U1} progressively executed the intermediate steps (Fig.~\ref{fig:case_ml}\captionID{B3} and \captionID{B4}) and discovered that the red rectangles appeared specifically in the instances generated by the repeating rule applied to the \textit{source} node. This indicated that while patterns where the heist node repeated 0 or 1 times produced valid results, those involving 2–3 repetitions did not yield any matches.

Using \techName{}, \textbf{U1} successfully expressed and verified her desired money laundering patterns within the dataset.

\begin{figure}[!htbp]
  \centering 
    \setlength{\belowcaptionskip}{-0.5cm}
  \includegraphics[width=0.95\linewidth
  ]{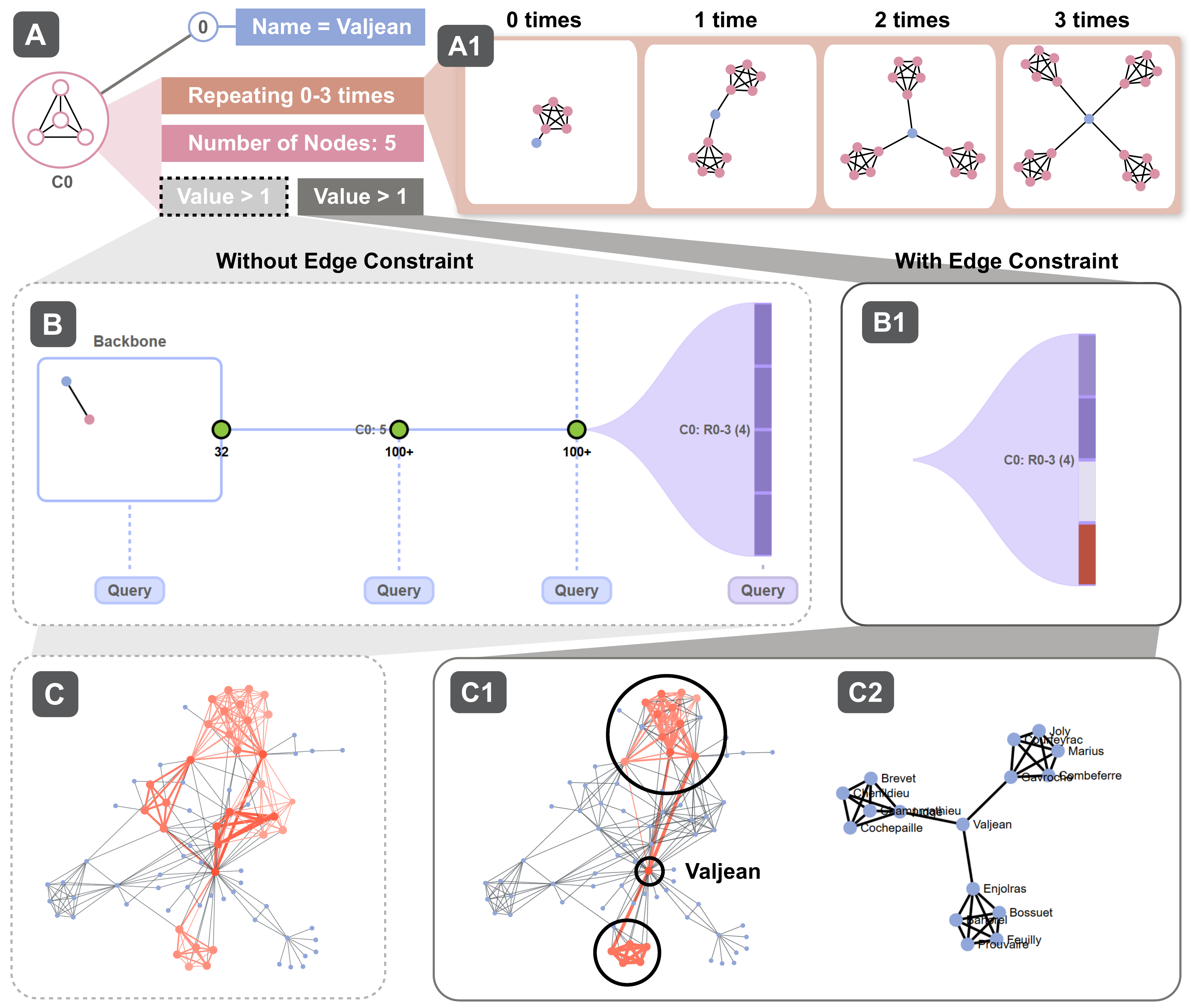}
  \caption{\modify{Community analysis in \textit{Les Mis{\'e}rables} using \techName{}. (A) The user queried \textit{Valjean}'s connections to multiple communities (A1).} (B) All query instances returned results, while applying the edge constraint led to fewer or no results for some instances (B1). By checking the highlighted results (C, C1, C2), the user identified two distinct communities connected to Valjean, corresponding to two major plotlines in the novel.}
  \label{fig:case_misarable}
\end{figure}

\subsection{Case 2: Rapid Character Relationship Querying}
\textbf{U9} has read the novel \textit{Les Mis{\'e}rables} and wanted to explore the relationships between its main characters by querying the character co-occurrence network (LMCN).

\textbf{Efficient Query Construction:} 
\textbf{U9} aimed to explore the relationships between the protagonist, \textit{Valjean}, and various communities (e.g., families or friend groups) by constructing a query, as shown in Fig.~\ref{fig:case_misarable}\captionID{A}. He began by adding \textit{Node 0} and applying a node attribute rule, ``\textit{name=Valjean}''. He then connected this node to a 5-node clique motif (\textit{C0}), assuming five members as a reasonable size for a community. \modify{To express multiple communities, \textbf{U9} intended to repeat the clique motif but was unsure how many such groups that \textit{Valjean} would be associated with, 
so he configured \textit{C0} to repeat 0-3 times. After reviewing the instantiation visualization (Fig.~\ref{fig:case_misarable}\captionID{A1}), \textbf{U9} confirmed that \techName{} correctly generated the query instances involving one to four communities connected to Valjean.}
Upon executing these instances through the \textit{Query Execution Panel}, \textbf{U9} observed that all query instances returned results (the green circles and purple blocks), indicating that \textit{Valjean} co-occurred with at least four distinct communities, each consisting of five or more characters. 
The characters in these communities are highlighted in the \textit{Result Overview}, as shown in Fig.~\ref{fig:case_misarable}\captionID{C}.

\textbf{Insightful Result Comparison:} 
\textbf{U9} considered that character co-occurrences happening only once might be incidental, potentially leading to the inaccurate identification of communities. 
To address this concern, he applied an edge attribute rule, ``\textit{value>1}'', to the clique motif (\textit{C0}), thereby constraining all edges within \textit{C0} to reflect stronger and more meaningful co-occurrence relationships of communities. 
After re-executing the query instances with this constraint, \textbf{U9} observed that the results differed from those generated without the edge value rule (Fig.~\ref{fig:case_misarable}\captionID{B1}), highlighting how edge strength influences community detection.
Specifically, query instances where \textit{Valjean} was connected to one to three communities still returned valid results, whereas the instance involving four communities yielded no results, as indicated by the red rectangles in Fig.~\ref{fig:case_misarable}\captionID{B1}.
The rectangle representing Valjean with three communities was colored light purple, suggesting a low number of matching results. 
Upon examining the \textit{Result Overview} (Fig.~\ref{fig:case_misarable}\captionID{C1}) and the \textit{Result List} (Fig.~\ref{fig:case_misarable}\captionID{C2}), \textbf{U9} found that although three communities were returned, they belonged to only two distinct communities, as shown by the two circled clusters in Fig.\ref{fig:case_misarable}\captionID{C1}.
By hovering over the nodes to view their names, \textbf{U9} discovered that the cluster in the upper circle primarily consisted of members of ``\textit{The Friends of the ABC}'', a revolutionary group that once fought alongside Valjean. The cluster in the lower circle included key characters involved in a subplot of an innocent man being mistaken for \textit{Valjean}.

\modify{With Envisage, \textbf{U9} quickly expressed and executed desired queries to explore \textit{Valjean}'s community relationships and intuitively uncover key insights, including two main plotlines in the novel.}

\section{User Interview}\label{sec: interview}
To evaluate the effectiveness of \techName{}, we conducted in-depth user interviews with 14 graph analysts. This section presents the participants, procedures, and key findings.

\subsection{Participants and Apparatus}

We recruited 14 participants (U1-U14) from universities for user interviews (6 females, 8 males, $age_{{mean}} = 24$, $age_{{sd}} = 2.95$, with normal vision and no color vision deficiency). 
All participants were majoring in fields related to computer science and had at least six months of experience in graph analysis. 
They comprised ten graduate students (U1-U3, U5-U8, and U12-U14), three undergraduates (U9-U11), and one postdoctoral researcher (U4).
All participants had prior experience writing code to extract information from graph data. However, only five participants (U5, U9–U12) were familiar with data querying languages or tools: U9 with Cypher, U10 and U11 with SQL, and U5 and U12 with NetworkX~\cite{networkx} and Cytoscape~\cite{cytoscape}.
All interviews were conducted online via Zoom. \techName{} was deployed on a remote server, and participants accessed it using their own laptops or desktops while sharing their screens with the interviewer.
Each interview lasted about one hour, and we paid \$15 to each participant for compensation.




\subsection{Procedures}
\begin{figure*}[!htbp]
  \centering 
  \setlength{\belowcaptionskip}{-0.5cm}
  \includegraphics[width=0.75\linewidth
  ]{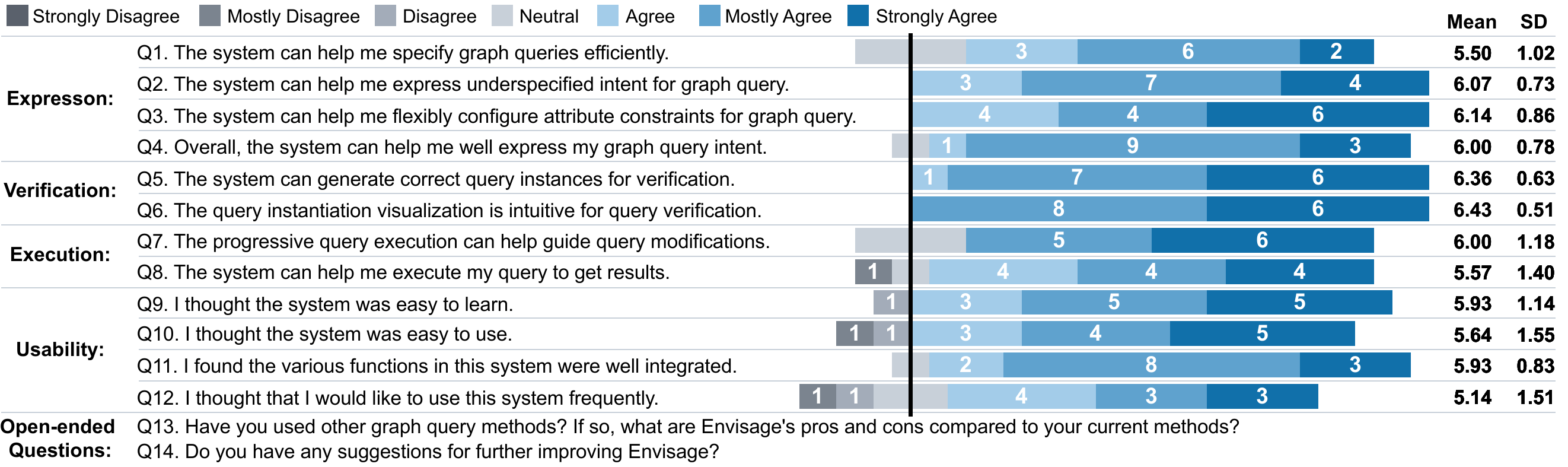}
  \caption{The user interview questionnaire results. Q1-Q12 are closed-ended questions rated on a 7-point Likert scale. Q13, Q14 are open-ended questions to collect participants’ feedback. The detailed scores of Q1-Q12 are shown in a stacked bar chart.}
  \label{fig:interview}
\end{figure*}

The user interviews consisted of tutorial, task, and interview phases.
In the tutorial phase, participants were first asked to access the online \techName{} system, and we then introduced its background, workflow, visual design, and interactions.
A usage scenario illustrated how \techName{} supports graph querying in practice.
We grouped participants into three groups based on their expertise and preferences, each working with a different dataset described in Section~\ref{sec: casestudy}.
\modify{In the task phase, we first introduced the assigned datasets and then guided participants through a series of instructions to perform graph querying. These instructions were designed to ensure that participants understood and used the core functions of \techName{}.} 
Afterward, participants were allowed to freely query specific patterns of their interest following our four-stage framework until they gained a comprehensive understanding of how \techName{} works.
The entire task phase typically lasted about 30 minutes.
Finally, participants completed a post-study questionnaire, which included 14 questions (Q1-Q14), as shown in Fig.~\ref{fig:interview}. 
Q1-Q12 were closed-ended, rated via a 7-point Likert scale~\cite{joshi2015likert} (1 for strongly disagree and 7 for strongly agree), and assessed \techName{} support for query expression (Q1-Q4), verification (Q5, Q6), execution (Q7, Q8), and usability (Q9-Q12). 
The usability questions (Q9-Q12) were selected from System Usability Scale (SUS) Questionnaire ~\cite{brooke2013sus}.

\subsection{Results}
Fig.~\ref{fig:interview} shows participants' feedback for closed-ended questions (Q1-Q12), including the score distribution along with the calculated mean (M) and standard deviation (SD) for each question. 
The questionnaire results indicate overall positive user feedback for \techName{} across key dimensions, while there was some negative feedback, which will be discussed in the detailed result analysis below.

\textbf{Expression:} 
Participants responded positively to \techName{}’s ability to support expressive and flexible graph query construction. \modify{They agreed it helped specify queries efficiently (Q1: M = 5.50) and express underspecified intent (Q2: M = 6.07), with high ratings for configuring attribute constraints (Q3: M = 6.14) and overall intent expression (Q4: M = 6.00).} 
However, some participants gave neutral responses, particularly to Q1 and Q4, primarily due to limitations in representing certain complex structures. 
For instance, U6 attempted to express two cliques connected by multiple edges across different nodes, but \techName{} treated inter-clique connections as linking to the same nodes. This simplification was necessary to prevent exponential growth in the number of generated query graphs as the node count increased. 
Similarly, U8 wanted to construct more complex or irregular query structures beyond the current support for repeating and chaining rules. Moving forward, these limitations could be addressed by enabling users to fine-tune generated instances or by translating query intent into graph matching programs rather than relying solely on enumerated query graphs and query languages.

\textbf{Verification \& Execution: }
\techName{} was highly rated for its support for verification, with users valuing the correctness of generated query instances (Q5: M = 6.36, SD = 0.63) and the clarity of query instantiation visualizations (Q6: M = 6.43, SD = 0.51). For execution, users appreciated the progressive execution feature, which helped guide query modifications (Q7: M = 6.00, SD = 1.18) and overall support for execution (Q8: M = 5.57, SD = 1.40).
Some users gave neutral or negative feedback on Q7 and Q8 due to slower response times when processing a large number of complex query instances, which impacted interaction smoothness. 
To align with \modify{underspecified} query intent, the system generates many possible query instances, placing heavy demands on computational resources, particularly memory. 
This issue could be alleviated by deploying the system on more powerful servers or incorporating techniques such as graph neural networks~\cite{hamilton2017graphsage} to accelerate querying rather than relying solely on database queries.

\textbf{Usability: }
Overall, participants found \techName{} easy to learn (Q9: M = 5.93) and use (Q10: M = 5.64), with well-integrated features (Q11: M = 5.93).
However, some neutral and negative feedback revealed areas for improvement. U7 noted the need for more detailed guidance, results statistics, and visual indicators during query processing, while U2 noted that repetitive basic interactions like adding nodes and edges can be simplified. The slightly lower score for intended frequent use (Q12: M = 5.14) was attributed to difficulties integrating \techName{} into existing workflows, which often involve additional steps such as deriving new attributes or converting outputs into domain-specific formats.

\textbf{Open-ended Questions:}
In Q13, participants compared \techName{} with other graph query methods. Those unfamiliar with such tools shared impressions based on a brief introduction of existing systems during user interviews. Overall, participants found \techName{} more expressive than traditional visual query tools, with features like query expansion rules helping them quickly express desired patterns. Participants familiar with graph query languages (U9 and U12) noted that these languages still offer greater expressive power, particularly for advanced operations such as subqueries, unions, and intersections. However, most participants noted that writing such queries remains time-consuming and laborious, even for experienced users. As a result, \techName{} is well-suited for querying complex graphs involving specific (e.g., motifs) and regular patterns (e.g., repeating structures), especially for novice users.
In response to Q14, U2 and U7 suggested incorporating a feature that allows users to describe their query intent in natural language. U2 further noted that while some recent studies have used large language models (LLMs) to generate Cypher queries~\cite{hornsteiner2024real}, these approaches remain limited in terms of query complexity, generation accuracy, and intuitive visual design. Therefore, combining LLMs with visual graph querying could be a promising direction for future exploration.
U12 also suggested making motif configuration more flexible, such as allowing customization of individual elements within a motif and enabling users to fine-tune query instances after instantiation.

\section{Discussion}
This section discusses key lessons learned about the expressiveness of visual graph querying and outlines our limitations.

\subsection{Expressiveness of Visual Graph Querying}
A primary goal of visual graph querying is to enable users to interactively construct graph queries that reflect their intent without relying on complex code or formal query languages~\cite{bhowmick2017graph}. However, user intent is often diverse and loosely defined. Existing approaches, such as \modify{``query-by-template''}, offer limited support for expressing more complex or underspecified queries. 
This work investigates how visual graph querying can support flexible query expression. 
Specifically, users can apply multiple rules to customized entities to construct complex query graphs aligned with their goals. 
Our prototype incorporates four common motifs and two additional parameterized rules (repeating and chaining) to assess the feasibility of this approach. 
Future extensions could support more customized motifs and rule types for broader expressiveness. 
User interviews revealed that \techName{} could express the most user-desired patterns, though minor gaps remained. 
One potential solution is to allow users to sketch rough patterns with current functions and fine-tune them as needed. 
Additionally, while prior work has used large language models (LLMs) for graph query generation~\cite{hornsteiner2024real}, these approaches struggle with complex patterns and are challenging to revise for users unfamiliar with query languages.
Combining the intent-capturing ability of LLMs with the visual expressiveness of systems like \techName{} presents a promising direction for more accessible and flexible graph querying.




\subsection{Limitations}
\modify{
\techName{} is not without limitations:
First, user queries are represented as subgraphs with attribute constraints, which can be translated into various query languages by extending the translators. However, both translation and execution performance may vary across languages. For example, Cypher queries can become verbose and slow when expressing large patterns, potentially impacting execution speed. Some languages like Gremlin~\cite{rodriguez2015gremlin} offer more efficient traversal constructs, but this comes at the cost of increased translator complexity.
Second, \techName{} can benefit from integrating existing techniques such as query auto-completion~\cite{yi2017autog} and auto-suggestion~\cite{jayaram2015viiq}. 
}
\modify{Third, the Result List can become difficult to navigate when a large number of results are returned, which could be mitigated by incorporating a dropdown menu or pagination.
Additionally, \techName{} could incorporate additional interactions to support the expression of temporal patterns. For example, in MLN dataset, edges represent timestamped transactions, and one participant aimed to specify temporal relationships between edges.}

\section{Conclusion}

We presented \techName{}, an interactive visual graph querying system designed to improve expressive power in query construction. \modify{\techName{} adopts a four-stage framework that guides users through the processes of expressing, verifying, and executing graph queries.} To evaluate it, we conducted two case studies and in-depth interviews with 14 graph analysts. The results show that \techName{} effectively supports users in building, refining, and executing complex queries with improved scalability and flexibility in handling \modify{underspecified query intent}.

In future work, we plan to allow users to customize rules when expressing graph queries, enhancing \techName{}’s adaptability across domains. We also aim to integrate natural language with visual interaction, allowing users to express query intent in natural language and refine generated query instances through an interactive visual interface.



\acknowledgments{%
This project is supported by the Ministry of Education, Singapore, under its Academic Research Fund Tier 2 (Proposal ID: T2EP20222-0049), and NTU Start Up Grant awarded to Yong Wang.
}

\bibliographystyle{abbrv-doi-hyperref}

\bibliography{main}

\begin{thebibliography}{10}

\bibitem{aberger2016emptyheaded}
C.~R. Aberger, S.~Tu, K.~Olukotun, and C.~Ré.
\newblock Emptyheaded: A relational engine for graph processing.
\newblock In {\em Proceedings of the International Conference on Management of Data}, pp. 431--446. ACM, 2016. \href{https://doi.org/10.1145/2882903.2915213}
{doi: {{%
10\hspace{.1pt}\discretionary{.}{%
}{.}\hspace{.4pt}1145\discretionary{/}{%
}{/}2882903\hspace{.1pt}\discretionary{.}{%
}{.}\hspace{.4pt}2915213}}}


\bibitem{amazon_neptune}
{Amazon Web Services, Inc.}
\newblock Amazon neptune: Fully managed graph database service.
\newblock \url{https://aws.amazon.com/neptune/}.
\newblock [Online; accessed 24-March-2025].

\bibitem{ammar2018bigjoin}
K.~Ammar, F.~McSherry, S.~Salihoglu, and M.~Joglekar.
\newblock Distributed evaluation of subgraph queries using {Worst‐Case} optimal low‐memory dataflows.
\newblock {\em arXiv preprint}, Feb. 2018. \href{https://doi.org/10.48550/arXiv.1802.03760}
{doi: {{%
10\hspace{.1pt}\discretionary{.}{%
}{.}\hspace{.4pt}48550\discretionary{/}{%
}{/}arXiv\hspace{.1pt}\discretionary{.}{%
}{.}\hspace{.4pt}1802\hspace{.1pt}\discretionary{.}{%
}{.}\hspace{.4pt}03760}}}


\bibitem{arangodb}
{ArangoDB GmbH}.
\newblock Arangodb: Multi-model database for your modern apps.
\newblock \url{https://arangodb.com/}.
\newblock [Online; accessed 24-March-2025].

\bibitem{bhowmick2022data}
S.~S. Bhowmick and B.~Choi.
\newblock Data-driven visual query interfaces for graphs: Past, present, and (near) future.
\newblock In {\em Proceedings of the International Conference on Management of Data}, pp. 2441--2447. ACM, Philadelphia, PA, USA, June 2022. \href{https://doi.org/10.1145/3514221.3522562}
{doi: {{%
10\hspace{.1pt}\discretionary{.}{%
}{.}\hspace{.4pt}1145\discretionary{/}{%
}{/}3514221\hspace{.1pt}\discretionary{.}{%
}{.}\hspace{.4pt}3522562}}}


\bibitem{bhowmick2017graph}
S.~S. Bhowmick, B.~Choi, and C.~Li.
\newblock Graph querying meets hci: State of the art and future directions.
\newblock In {\em Proceedings of the International Conference on Management of Data}, pp. 1731--1736. ACM, Chicago, May 2017. \href{https://doi.org/10.1145/3035918.3054774}
{doi: {{%
10\hspace{.1pt}\discretionary{.}{%
}{.}\hspace{.4pt}1145\discretionary{/}{%
}{/}3035918\hspace{.1pt}\discretionary{.}{%
}{.}\hspace{.4pt}3054774}}}


\bibitem{bhowmick2013vogue}
S.~S. Bhowmick, B.~Choi, and S.~Zhou.
\newblock {VOGUE}: Towards a visual interaction-aware graph query processing framework.
\newblock In {\em Proceedings of the Conference on Innovative Data Systems Research}, 2013.

\bibitem{bhowmick2015visual}
S.~S. Bhowmick, H.~E. Chua, B.~Thian, and B.~Choi.
\newblock {ViSual}: An hci-inspired simulator for blending visual subgraph query construction and processing.
\newblock In {\em Proceedings of the 31st International Conference on Data Engineering}, pp. 1480--1483. IEEE, Seoul, Apr. 2015. \href{https://doi.org/10.1109/ICDE.2015.7113406}
{doi: {{%
10\hspace{.1pt}\discretionary{.}{%
}{.}\hspace{.4pt}1109\discretionary{/}{%
}{/}ICDE\hspace{.1pt}\discretionary{.}{%
}{.}\hspace{.4pt}2015\hspace{.1pt}\discretionary{.}{%
}{.}\hspace{.4pt}7113406}}}


\bibitem{brooke2013sus}
J.~Brooke.
\newblock {SUS}: A retrospective.
\newblock {\em Journal of usability studies}, 8(2):29--40, 2013. \href{https://dl.acm.org/doi/abs/10.5555/2817912.2817913}
{doi: {{%
doi\discretionary{/}{%
}{/}abs\discretionary{/}{%
}{/}10\hspace{.1pt}\discretionary{.}{%
}{.}\hspace{.4pt}5555\discretionary{/}{%
}{/}2817912\hspace{.1pt}\discretionary{.}{%
}{.}\hspace{.4pt}2817913}}}


\bibitem{cakmak2022motif-based}
E.~Cakmak, J.~Fuchs, D.~Jäckle, T.~Schreck, U.~Brandes, and D.~Keim.
\newblock Motif-based visual analysis of dynamic networks.
\newblock In {\em Proceedings of the Visualization in Data Science}, pp. 17--26. IEEE, Oct. 2022. \href{https://doi.org/10.1109/VDS57266.2022.00007}
{doi: {{%
10\hspace{.1pt}\discretionary{.}{%
}{.}\hspace{.4pt}1109\discretionary{/}{%
}{/}VDS57266\hspace{.1pt}\discretionary{.}{%
}{.}\hspace{.4pt}2022\hspace{.1pt}\discretionary{.}{%
}{.}\hspace{.4pt}00007}}}


\bibitem{chau2008graphite}
D.~H. Chau, C.~Faloutsos, H.~Tong, J.~I. Hong, B.~Gallagher, and T.~Eliassi-Rad.
\newblock Graphite: A visual query system for large graphs.
\newblock In {\em Proceedings of the International Conference on Data Mining Workshops}, pp. 963--966. IEEE, 2008. \href{https://doi.org/10.1109/ICDMW.2008.99}
{doi: {{%
10\hspace{.1pt}\discretionary{.}{%
}{.}\hspace{.4pt}1109\discretionary{/}{%
}{/}ICDMW\hspace{.1pt}\discretionary{.}{%
}{.}\hspace{.4pt}2008\hspace{.1pt}\discretionary{.}{%
}{.}\hspace{.4pt}99}}}


\bibitem{cordella2004vf2}
L.~P. Cordella, P.~Foggia, C.~Sansone, and M.~Vento.
\newblock A (sub)graph isomorphism algorithm for matching large graphs.
\newblock {\em IEEE Transactions on Pattern Analysis and Machine Intelligence}, 26(10):1367--1372, Oct. 2004. \href{https://doi.org/10.1109/TPAMI.2004.75}
{doi: {{%
10\hspace{.1pt}\discretionary{.}{%
}{.}\hspace{.4pt}1109\discretionary{/}{%
}{/}TPAMI\hspace{.1pt}\discretionary{.}{%
}{.}\hspace{.4pt}2004\hspace{.1pt}\discretionary{.}{%
}{.}\hspace{.4pt}75}}}


\bibitem{cuenca2021vertigo}
E.~Cuenca, A.~Sallaberry, D.~Ienco, and P.~Poncelet.
\newblock Vertigo: A visual platform for querying and exploring large multilayer networks.
\newblock {\em IEEE Transactions on Visualization and Computer Graphics}, 28(3):1634--1647, 2021. \href{https://doi.org/10.1109/TVCG.2021.3067820}
{doi: {{%
10\hspace{.1pt}\discretionary{.}{%
}{.}\hspace{.4pt}1109\discretionary{/}{%
}{/}TVCG\hspace{.1pt}\discretionary{.}{%
}{.}\hspace{.4pt}2021\hspace{.1pt}\discretionary{.}{%
}{.}\hspace{.4pt}3067820}}}


\bibitem{cytoscape}
{Cytoscape Consortium}.
\newblock Cytoscape: An open source platform for complex network analysis and visualization.
\newblock \url{https://cytoscape.org/}, 2024.
\newblock [Online; accessed: 28-March-2025].

\bibitem{dunne2013motif}
C.~Dunne and B.~Shneiderman.
\newblock Motif simplification: Improving network visualization readability with fan, connector, and clique glyphs.
\newblock In {\em Proceedings of the CHI Conference on Human Factors in Computing Systems}, pp. 3247--3256. ACM, Paris, France, Apr. 2013. \href{https://doi.org/10.1145/2470654.2466444}
{doi: {{%
10\hspace{.1pt}\discretionary{.}{%
}{.}\hspace{.4pt}1145\discretionary{/}{%
}{/}2470654\hspace{.1pt}\discretionary{.}{%
}{.}\hspace{.4pt}2466444}}}


\bibitem{egi2018loop}
S.~Egi.
\newblock Loop patterns: Extension of kleene star operator for more expressive pattern matching against arbitrary data structures.
\newblock {\em arXiv preprint}, Sept. 2018. \href{https://doi.org/10.48550/arXiv.1809.03252}
{doi: {{%
10\hspace{.1pt}\discretionary{.}{%
}{.}\hspace{.4pt}48550\discretionary{/}{%
}{/}arXiv\hspace{.1pt}\discretionary{.}{%
}{.}\hspace{.4pt}1809\hspace{.1pt}\discretionary{.}{%
}{.}\hspace{.4pt}03252}}}


\bibitem{francis2018cypher}
N.~Francis, A.~Green, P.~Guagliardo, L.~Libkin, T.~Lindaaker, V.~Marsault, S.~Plantikow, M.~Rydberg, P.~Selmer, and A.~Taylor.
\newblock Cypher: An evolving query language for property graphs.
\newblock In {\em Proceedings of the 2018 International Conference on Management of Data}, pp. 1433--1445. ACM, 2018. \href{https://doi.org/10.1145/3183713.3190657}
{doi: {{%
10\hspace{.1pt}\discretionary{.}{%
}{.}\hspace{.4pt}1145\discretionary{/}{%
}{/}3183713\hspace{.1pt}\discretionary{.}{%
}{.}\hspace{.4pt}3190657}}}


\bibitem{freeman2004development}
L.~Freeman et~al.
\newblock The development of social network analysis.
\newblock {\em A Study in the Sociology of Science}, 1(687):159--167, 2004.

\bibitem{fuchs2024exploring}
J.~Fuchs, F.~L. Dennig, M.~Heinle, D.~A. Keim, and S.~Di~Bartolomeo.
\newblock Exploring the design space of biofabric visualization for multivariate network analysis.
\newblock {\em Computer Graphics Forum}, 43(3):e15079, June 2024. \href{https://doi.org/10.1111/cgf.15079}
{doi: {{%
10\hspace{.1pt}\discretionary{.}{%
}{.}\hspace{.4pt}1111\discretionary{/}{%
}{/}cgf\hspace{.1pt}\discretionary{.}{%
}{.}\hspace{.4pt}15079}}}


\bibitem{hamilton2017graphsage}
W.~Hamilton, Z.~Ying, and J.~Leskovec.
\newblock Inductive representation learning on large graphs.
\newblock {\em Advances in neural information processing systems}, 30:1025--1035, 2017. \href{https://doi.org/10.5555/3294771.3294869}
{doi: {{%
10\hspace{.1pt}\discretionary{.}{%
}{.}\hspace{.4pt}5555\discretionary{/}{%
}{/}3294771\hspace{.1pt}\discretionary{.}{%
}{.}\hspace{.4pt}3294869}}}


\bibitem{han2019daf}
M.~Han, H.~Kim, G.~Gu, K.~Park, and W.~Han.
\newblock Efficient subgraph matching: Harmonizing dynamic programming, adaptive matching order, and failing set together.
\newblock In {\em Proceedings of the International Conference on Management of Data}, pp. 1429--1440. ACM, Amsterdam, July 2019. \href{https://doi.org/10.1145/3299869.3319880}
{doi: {{%
10\hspace{.1pt}\discretionary{.}{%
}{.}\hspace{.4pt}1145\discretionary{/}{%
}{/}3299869\hspace{.1pt}\discretionary{.}{%
}{.}\hspace{.4pt}3319880}}}


\bibitem{han2013turboiso}
W.~Han, J.~Lee, and J.~Lee.
\newblock {TurboISO}: Towards ultrafast and robust subgraph isomorphism search in large graph databases.
\newblock In {\em Proceedings of the International Conference on Management of Data}, pp. 337--348. ACM, New York, June 2013. \href{https://doi.org/10.1145/2463676.2465300}
{doi: {{%
10\hspace{.1pt}\discretionary{.}{%
}{.}\hspace{.4pt}1145\discretionary{/}{%
}{/}2463676\hspace{.1pt}\discretionary{.}{%
}{.}\hspace{.4pt}2465300}}}


\bibitem{he2008graphql}
H.~He and A.~K. Singh.
\newblock Graphs-at-a-time: Query language and access methods for graph databases.
\newblock In {\em Proceedings of the International Conference on Management of Data}, pp. 405--418. ACM, Vancouver, June 2008. \href{https://doi.org/10.1145/1376616.1376660}
{doi: {{%
10\hspace{.1pt}\discretionary{.}{%
}{.}\hspace{.4pt}1145\discretionary{/}{%
}{/}1376616\hspace{.1pt}\discretionary{.}{%
}{.}\hspace{.4pt}1376660}}}


\bibitem{hornsteiner2024real}
M.~Hornsteiner, M.~Kreussel, C.~Steindl, F.~Ebner, P.~Empl, and S.~Sch{\"o}nig.
\newblock Real-time text-to-cypher query generation with large language models for graph databases.
\newblock {\em Future Internet}, 16(12):438, Dec. 2024. \href{https://doi.org/10.3390/fi16120438}
{doi: {{%
10\hspace{.1pt}\discretionary{.}{%
}{.}\hspace{.4pt}3390\discretionary{/}{%
}{/}fi16120438}}}


\bibitem{huang2023visualneo}
K.~Huang, H.~Liang, C.~Yao, X.~Zhao, Y.~Cui, Y.~Tian, R.~Zhang, and X.~Zhou.
\newblock Visualneo: Bridging the gap between visual query interfaces and graph query engines.
\newblock {\em Proceedings of the VLDB Endowment}, 16(12):4010--4013, Sept. 2023. \href{https://doi.org/10.14778/3611540.3611608}
{doi: {{%
10\hspace{.1pt}\discretionary{.}{%
}{.}\hspace{.4pt}14778\discretionary{/}{%
}{/}3611540\hspace{.1pt}\discretionary{.}{%
}{.}\hspace{.4pt}3611608}}}


\bibitem{huang05motif}
W.~Huang, C.~Murray, X.~Shen, L.~Song, Y.~X. Wu, and L.~Zheng.
\newblock Visualisation and analysis of network motifs.
\newblock In {\em Proceedings of the Ninth International Conference on Information Visualisation}, pp. 697--702, 2005. \href{https://doi.org/10.1109/IV.2005.138}
{doi: {{%
10\hspace{.1pt}\discretionary{.}{%
}{.}\hspace{.4pt}1109\discretionary{/}{%
}{/}IV\hspace{.1pt}\discretionary{.}{%
}{.}\hspace{.4pt}2005\hspace{.1pt}\discretionary{.}{%
}{.}\hspace{.4pt}138}}}


\bibitem{jayaram2015viiq}
N.~Jayaram, S.~Goyal, and C.~Li.
\newblock {VIIQ}: Auto-suggestion enabled visual interface for interactive graph query formulation.
\newblock {\em Proceedings of the VLDB Endowment}, 8(12):1940--1943, Aug. 2015. \href{https://doi.org/10.14778/2824032.2824106}
{doi: {{%
10\hspace{.1pt}\discretionary{.}{%
}{.}\hspace{.4pt}14778\discretionary{/}{%
}{/}2824032\hspace{.1pt}\discretionary{.}{%
}{.}\hspace{.4pt}2824106}}}


\bibitem{jin2012prague}
C.~Jin, S.~S. Bhowmick, B.~Choi, and S.~Zhou.
\newblock Prague: Towards blending practical visual subgraph query formulation and query processing.
\newblock In {\em Proceedings of the 28th International Conference on Data Engineering}, pp. 222--233. IEEE, Washington, Apr. 2012. \href{https://doi.org/10.1109/ICDE.2012.49}
{doi: {{%
10\hspace{.1pt}\discretionary{.}{%
}{.}\hspace{.4pt}1109\discretionary{/}{%
}{/}ICDE\hspace{.1pt}\discretionary{.}{%
}{.}\hspace{.4pt}2012\hspace{.1pt}\discretionary{.}{%
}{.}\hspace{.4pt}49}}}


\bibitem{joshi2015likert}
A.~Joshi, S.~Kale, S.~Chandel, and D.~K. Pal.
\newblock Likert scale: Explored and explained.
\newblock {\em British Journal of Applied Science \& Technology}, 7(4):396--403, 2015. \href{https://doi.org/10.9734/BJAST/2015/14975}
{doi: {{%
10\hspace{.1pt}\discretionary{.}{%
}{.}\hspace{.4pt}9734\discretionary{/}{%
}{/}BJAST\discretionary{/}{%
}{/}2015\discretionary{/}{%
}{/}14975}}}


\bibitem{jung2024monetexplorer}
S.~Jung, D.~Shin, H.~Jeon, K.~Choe, and J.~Seo.
\newblock {MoNetExplorer}: A visual analytics system for analyzing dynamic networks with temporal network motifs.
\newblock {\em IEEE Transactions on Visualization and Computer Graphics}, 30(10):6725--6739, Oct. 2024. \href{https://doi.org/10.1109/TVCG.2023.3337396}
{doi: {{%
10\hspace{.1pt}\discretionary{.}{%
}{.}\hspace{.4pt}1109\discretionary{/}{%
}{/}TVCG\hspace{.1pt}\discretionary{.}{%
}{.}\hspace{.4pt}2023\hspace{.1pt}\discretionary{.}{%
}{.}\hspace{.4pt}3337396}}}


\bibitem{juettner2018vf2plusplus}
A.~J{\"u}ttner and P.~Madarasi.
\newblock {VF2++}: An improved subgraph isomorphism algorithm.
\newblock {\em Discrete Applied Mathematics}, 242:69--81, June 2018.
\newblock Computational Advances in Combinatorial Optimization. \href{https://doi.org/10.1016/j.dam.2018.02.018}
{doi: {{%
10\hspace{.1pt}\discretionary{.}{%
}{.}\hspace{.4pt}1016\discretionary{/}{%
}{/}j\hspace{.1pt}\discretionary{.}{%
}{.}\hspace{.4pt}dam\hspace{.1pt}\discretionary{.}{%
}{.}\hspace{.4pt}2018\hspace{.1pt}\discretionary{.}{%
}{.}\hspace{.4pt}02\hspace{.1pt}\discretionary{.}{%
}{.}\hspace{.4pt}018}}}


\bibitem{knuth1993stanford}
D.~E. Knuth.
\newblock The stanford graphbase: A platform for combinatorial algorithms.
\newblock In {\em Proceedings of the 4th Annual ACM-SIAM Symposium on Discrete Algorithms}, pp. 41--43. ACM/SIAM, Austin, Jan. 1993. \href{https://dblp.org/rec/conf/soda/Knuth93}
{doi: {{%
rec\discretionary{/}{%
}{/}conf\discretionary{/}{%
}{/}soda\discretionary{/}{%
}{/}Knuth93}}}


\bibitem{lai2019distributed}
L.~Lai, Z.~Qing, Z.~Yang, X.~Jin, Z.~Lai, R.~Wang, K.~Hao, X.~Lin, L.~Qin, W.~Zhang, Y.~Zhang, Z.~Qian, and J.~Zhou.
\newblock Distributed subgraph matching on timely dataflow.
\newblock {\em Proceedings of the VLDB Endowment}, 12(10):1099--1112, 2019. \href{https://doi.org/10.14778/3339490.3339494}
{doi: {{%
10\hspace{.1pt}\discretionary{.}{%
}{.}\hspace{.4pt}14778\discretionary{/}{%
}{/}3339490\hspace{.1pt}\discretionary{.}{%
}{.}\hspace{.4pt}3339494}}}


\bibitem{li2022motif}
C.~Li, Y.~Tang, Z.~Tang, J.~Cao, and Y.~Zhang.
\newblock Motif-based embedding label propagation algorithm for community detection.
\newblock {\em International Journal of Intelligent Systems}, 37(3):1880--1902, 2022. \href{https://doi.org/10.1002/int.22759}
{doi: {{%
10\hspace{.1pt}\discretionary{.}{%
}{.}\hspace{.4pt}1002\discretionary{/}{%
}{/}int\hspace{.1pt}\discretionary{.}{%
}{.}\hspace{.4pt}22759}}}


\bibitem{li2024hiregex}
G.~Li, H.~Mi, C.~H. Liu, T.~Itoh, and G.~Wang.
\newblock {HiRegEx}: Interactive visual query and exploration of multivariate hierarchical data.
\newblock {\em IEEE Transactions on Visualization and Computer Graphics}, 2024. \href{https://doi.org/10.1109/TVCG.2024.3456389}
{doi: {{%
10\hspace{.1pt}\discretionary{.}{%
}{.}\hspace{.4pt}1109\discretionary{/}{%
}{/}TVCG\hspace{.1pt}\discretionary{.}{%
}{.}\hspace{.4pt}2024\hspace{.1pt}\discretionary{.}{%
}{.}\hspace{.4pt}3456389}}}


\bibitem{lin2024denseflow}
D.~Lin, J.~Wu, Y.~Yu, Q.~Fu, Z.~Zheng, and C.~Yang.
\newblock Denseflow: Spotting cryptocurrency money laundering in ethereum transaction graphs.
\newblock In {\em Proceedings of the ACM on Web Conference 2024}, pp. 4429--4438, 2024. \href{https://doi.org/10.1145/3589334.3645692}
{doi: {{%
10\hspace{.1pt}\discretionary{.}{%
}{.}\hspace{.4pt}1145\discretionary{/}{%
}{/}3589334\hspace{.1pt}\discretionary{.}{%
}{.}\hspace{.4pt}3645692}}}


\bibitem{ma2024sierra}
J.~Ma, S.~S. Bhowmick, L.~Tay, and B.~Choi.
\newblock {SIERRA}: A counterfactual thinking-based visual interface for property graph query construction.
\newblock In {\em Companion of the 2024 International Conference on Management of Data}, pp. 440--443. ACM, New York, June 2024. \href{https://doi.org/10.1145/3626246.3654729}
{doi: {{%
10\hspace{.1pt}\discretionary{.}{%
}{.}\hspace{.4pt}1145\discretionary{/}{%
}{/}3626246\hspace{.1pt}\discretionary{.}{%
}{.}\hspace{.4pt}3654729}}}


\bibitem{mhedhbi2019optimizing}
A.~Mhedhbi and S.~Salihoglu.
\newblock Optimizing subgraph queries by combining binary and worst-case optimal joins.
\newblock {\em Proceedings of the VLDB Endowment}, 12(11):1692--1704, 2019. \href{https://doi.org/10.14778/3342263.3342643}
{doi: {{%
10\hspace{.1pt}\discretionary{.}{%
}{.}\hspace{.4pt}14778\discretionary{/}{%
}{/}3342263\hspace{.1pt}\discretionary{.}{%
}{.}\hspace{.4pt}3342643}}}


\bibitem{miller2013graph}
J.~J. Miller.
\newblock Graph database applications and concepts with neo4j.
\newblock In {\em Proceedings of the Southern Association for Information Systems Conference}, pp. 141--147. Atlanta, 2013.

\bibitem{neo4j}
{Neo4j, Inc.}
\newblock Neo4j graph database \& analytics.
\newblock \url{https://neo4j.com/}.
\newblock [Online; accessed 24-March-2025].

\bibitem{networkx}
{NetworkX Developers}.
\newblock {NetworkX}: Network analysis in python.
\newblock \url{https://networkx.org/}, 2024.
\newblock [Online; accessed: 28-March-2025].

\bibitem{perez2009sparql}
J.~P{\'e}rez, M.~Arenas, and C.~Gutierrez.
\newblock Semantics and complexity of sparql.
\newblock {\em ACM Transactions on Database Systems}, 34(3):16:1--16:45, 2009. \href{https://doi.org/10.1145/1567274.1567278}
{doi: {{%
10\hspace{.1pt}\discretionary{.}{%
}{.}\hspace{.4pt}1145\discretionary{/}{%
}{/}1567274\hspace{.1pt}\discretionary{.}{%
}{.}\hspace{.4pt}1567278}}}


\bibitem{pienta2017vigor}
R.~Pienta, F.~Hohman, A.~Endert, A.~Tamersoy, K.~Roundy, C.~Gates, S.~Navathe, and D.~H. Chau.
\newblock Vigor: Interactive visual exploration of graph query results.
\newblock {\em IEEE Transactions on Visualization and Computer Graphics}, 24(1):215--225, 2017. \href{https://doi.org/10.1109/TVCG.2017.2744898}
{doi: {{%
10\hspace{.1pt}\discretionary{.}{%
}{.}\hspace{.4pt}1109\discretionary{/}{%
}{/}TVCG\hspace{.1pt}\discretionary{.}{%
}{.}\hspace{.4pt}2017\hspace{.1pt}\discretionary{.}{%
}{.}\hspace{.4pt}2744898}}}


\bibitem{pienta2016visage}
R.~Pienta, A.~Tamersoy, A.~Endert, S.~Navathe, H.~Tong, and D.~H. Chau.
\newblock Visage: Interactive visual graph querying.
\newblock In {\em Proceedings of the International Working Conference on Advanced Visual Interfaces}, pp. 272--279, 2016. \href{https://doi.org/10.1145/2909132.2909246}
{doi: {{%
10\hspace{.1pt}\discretionary{.}{%
}{.}\hspace{.4pt}1145\discretionary{/}{%
}{/}2909132\hspace{.1pt}\discretionary{.}{%
}{.}\hspace{.4pt}2909246}}}


\bibitem{pister2022visual}
A.~Pister, C.~Prieur, and J.-D. Fekete.
\newblock Visual queries on bipartite multivariate dynamic social networks.
\newblock In {\em Proceedings of the 24th Eurographics Conference on Visualization}, 2022. \href{https://doi.org/10.2312/evp.20221115}
{doi: {{%
10\hspace{.1pt}\discretionary{.}{%
}{.}\hspace{.4pt}2312\discretionary{/}{%
}{/}evp\hspace{.1pt}\discretionary{.}{%
}{.}\hspace{.4pt}20221115}}}


\bibitem{ren2015exploiting}
X.~Ren and J.~Wang.
\newblock Exploiting vertex relationships in speeding up subgraph isomorphism over large graphs.
\newblock {\em Proceedings of the VLDB Endowment}, 8(5):617--628, Jan. 2015. \href{https://doi.org/10.14778/2735479.2735493}
{doi: {{%
10\hspace{.1pt}\discretionary{.}{%
}{.}\hspace{.4pt}14778\discretionary{/}{%
}{/}2735479\hspace{.1pt}\discretionary{.}{%
}{.}\hspace{.4pt}2735493}}}


\bibitem{rivero2017efficient}
C.~R. Rivero and H.~M. Jamil.
\newblock Efficient and scalable labeled subgraph matching using sgmatch.
\newblock {\em Knowledge and Information Systems}, 51(1):61--87, 2017. \href{https://doi.org/10.1007/s10115-016-0968-2}
{doi: {{%
10\hspace{.1pt}\discretionary{.}{%
}{.}\hspace{.4pt}1007\discretionary{/}{%
}{/}s10115\discretionary{%
}{-}{-}016\discretionary{%
}{-}{-}0968\discretionary{%
}{-}{-}2}}}


\bibitem{halborn2021explained}
{Rob Behnke}.
\newblock {Explained: The EasyFi Hack April 2021}.
\newblock \url{https://www.halborn.com/blog/post/explained-the-easyfi-hack-april-2021}, 2021.
\newblock [Online; accessed 19-March-2025].

\bibitem{rodriguez2015gremlin}
M.~A. Rodriguez.
\newblock The gremlin graph traversal machine and language.
\newblock In {\em Proceedings of the 15th Symposium on Database Programming Languages}, pp. 1--10. ACM, Kohala Coast, Aug. 2015. \href{https://doi.org/10.1145/2815072.2815073}
{doi: {{%
10\hspace{.1pt}\discretionary{.}{%
}{.}\hspace{.4pt}1145\discretionary{/}{%
}{/}2815072\hspace{.1pt}\discretionary{.}{%
}{.}\hspace{.4pt}2815073}}}


\bibitem{shang2008taming}
H.~Shang, Y.~Zhang, X.~Lin, and J.~X. Yu.
\newblock Taming verification hardness: An efficient algorithm for testing subgraph isomorphism.
\newblock {\em Proceedings of the VLDB Endowment}, 1(1):364--375,  12 pages, Aug. 2008. \href{https://doi.org/10.14778/1453856.1453899}
{doi: {{%
10\hspace{.1pt}\discretionary{.}{%
}{.}\hspace{.4pt}14778\discretionary{/}{%
}{/}1453856\hspace{.1pt}\discretionary{.}{%
}{.}\hspace{.4pt}1453899}}}


\bibitem{song2021interactive}
H.~Song, Z.~Dai, P.~Xu, and L.~Ren.
\newblock Interactive visual pattern search on graph data via graph representation learning.
\newblock {\em IEEE Transactions on Visualization and Computer Graphics}, 28(1):335--345, 2021. \href{https://doi.org/10.1109/TVCG.2021.3114857}
{doi: {{%
10\hspace{.1pt}\discretionary{.}{%
}{.}\hspace{.4pt}1109\discretionary{/}{%
}{/}TVCG\hspace{.1pt}\discretionary{.}{%
}{.}\hspace{.4pt}2021\hspace{.1pt}\discretionary{.}{%
}{.}\hspace{.4pt}3114857}}}


\bibitem{sun2021rapidmatch}
S.~Sun, X.~Sun, Y.~Che, Q.~Luo, and B.~He.
\newblock Rapidmatch: A holistic approach to subgraph query processing.
\newblock {\em Proceedings of the VLDB Endowment}, 14(2):176--188, 2021. \href{https://doi.org/10.14778/3425879.3425888}
{doi: {{%
10\hspace{.1pt}\discretionary{.}{%
}{.}\hspace{.4pt}14778\discretionary{/}{%
}{/}3425879\hspace{.1pt}\discretionary{.}{%
}{.}\hspace{.4pt}3425888}}}


\bibitem{tamersoy2014large}
A.~Tamersoy, E.~Khalil, B.~Xie, S.~L. Lenkey, B.~R. Routledge, D.~H. Chau, and S.~B. Navathe.
\newblock Large-scale insider trading analysis: patterns and discoveries.
\newblock {\em Social Network Analysis and Mining}, 4:1--17, 2014. \href{https://doi.org/10.1007/s13278-014-0201-9}
{doi: {{%
10\hspace{.1pt}\discretionary{.}{%
}{.}\hspace{.4pt}1007\discretionary{/}{%
}{/}s13278\discretionary{%
}{-}{-}014\discretionary{%
}{-}{-}0201\discretionary{%
}{-}{-}9}}}


\bibitem{thompson1968programming}
K.~Thompson.
\newblock Programming techniques: Regular expression search algorithm.
\newblock {\em Communications of the ACM}, 11(6):419--422, June 1968. \href{https://doi.org/10.1145/363347.363387}
{doi: {{%
10\hspace{.1pt}\discretionary{.}{%
}{.}\hspace{.4pt}1145\discretionary{/}{%
}{/}363347\hspace{.1pt}\discretionary{.}{%
}{.}\hspace{.4pt}363387}}}


\bibitem{troidl2023vimo}
J.~Troidl, S.~Warchol, J.~Choi, J.~Matelsky, N.~Dhanyasi, X.~Wang, B.~Wester, D.~Wei, J.~W. Lichtman, and H.~Pfister.
\newblock Vimo – visual analysis of neuronal connectivity motifs.
\newblock {\em IEEE Transactions on Visualization and Computer Graphics}, 30(1):748--758, 2023. \href{https://doi.org/10.1109/TVCG.2023.3327388}
{doi: {{%
10\hspace{.1pt}\discretionary{.}{%
}{.}\hspace{.4pt}1109\discretionary{/}{%
}{/}TVCG\hspace{.1pt}\discretionary{.}{%
}{.}\hspace{.4pt}2023\hspace{.1pt}\discretionary{.}{%
}{.}\hspace{.4pt}3327388}}}


\bibitem{van2013dynamic}
S.~Van Den~Elzen, D.~Holten, J.~Blaas, and J.~J. Van~Wijk.
\newblock Dynamic network visualization with extended massive sequence views.
\newblock {\em IEEE Transactions on Visualization and Computer Graphics}, 20(8):1087--1099, Nov. 2013. \href{https://doi.org/10.1109/TVCG.2013.263}
{doi: {{%
10\hspace{.1pt}\discretionary{.}{%
}{.}\hspace{.4pt}1109\discretionary{/}{%
}{/}TVCG\hspace{.1pt}\discretionary{.}{%
}{.}\hspace{.4pt}2013\hspace{.1pt}\discretionary{.}{%
}{.}\hspace{.4pt}263}}}


\bibitem{vargas2020user}
H.~Vargas, C.~Buil-Aranda, A.~Hogan, and C.~L\'{o}pez.
\newblock A user interface for exploring and querying knowledge graphs (extended abstract).
\newblock In {\em Proceedings of the 29th International Joint Conference on Artificial Intelligence},  article no. 666,  5 pages, 2021. \href{https://dl.acm.org/doi/abs/10.5555/3491440.3492106}
{doi: {{%
doi\discretionary{/}{%
}{/}abs\discretionary{/}{%
}{/}10\hspace{.1pt}\discretionary{.}{%
}{.}\hspace{.4pt}5555\discretionary{/}{%
}{/}3491440\hspace{.1pt}\discretionary{.}{%
}{.}\hspace{.4pt}3492106}}}


\bibitem{von2009visual}
T.~von Landesberger, M.~Gorner, and T.~Schreck.
\newblock Visual analysis of graphs with multiple connected components.
\newblock In {\em Proceedings of the Symposium on Visual Analytics Science and Technology}, pp. 155--162. IEEE, Oct. 2009. \href{https://doi.org/10.1109/VAST.2009.5333893}
{doi: {{%
10\hspace{.1pt}\discretionary{.}{%
}{.}\hspace{.4pt}1109\discretionary{/}{%
}{/}VAST\hspace{.1pt}\discretionary{.}{%
}{.}\hspace{.4pt}2009\hspace{.1pt}\discretionary{.}{%
}{.}\hspace{.4pt}5333893}}}


\bibitem{wen2023nftdisk}
X.~Wen, Y.~Wang, X.~Yue, F.~Zhu, and M.~Zhu.
\newblock {NFTDisk}: Visual detection of wash trading in nft markets.
\newblock In {\em Proceedings of the CHI Conference on Human Factors in Computing Systems}, pp. 1--15. ACM, 2023. \href{https://doi.org/10.1145/3544548.3581466}
{doi: {{%
10\hspace{.1pt}\discretionary{.}{%
}{.}\hspace{.4pt}1145\discretionary{/}{%
}{/}3544548\hspace{.1pt}\discretionary{.}{%
}{.}\hspace{.4pt}3581466}}}


\bibitem{WikipediaLesMis}
{Wikipedia contributors}.
\newblock {\em Les Mis{\'e}rables} --- wikipedia.
\newblock \url{https://en.wikipedia.org/w/index.php?title=Les_Mis\%C3\%A9rables\&oldid=1006098025}, Mar. 2025.
\newblock [Online; accessed 19-March-2025].

\bibitem{wu2023toward}
J.~Wu, D.~Lin, Q.~Fu, S.~Yang, T.~Chen, Z.~Zheng, and B.~Song.
\newblock Toward understanding asset flows in crypto money laundering through the lenses of {Ethereum} heists.
\newblock {\em IEEE Transactions on Information Forensics and Security}, 19:1994--2009, 2023. \href{https://doi.org/10.1109/TIFS.2023.3346276}
{doi: {{%
10\hspace{.1pt}\discretionary{.}{%
}{.}\hspace{.4pt}1109\discretionary{/}{%
}{/}TIFS\hspace{.1pt}\discretionary{.}{%
}{.}\hspace{.4pt}2023\hspace{.1pt}\discretionary{.}{%
}{.}\hspace{.4pt}3346276}}}


\bibitem{yang2014schemaless}
S.~Yang, Y.~Wu, H.~Sun, and X.~Yan.
\newblock Schemaless and structureless graph querying.
\newblock {\em Proceedings of the VLDB Endowment}, 7(7):565--576, 2014. \href{https://doi.org/10.14778/2732286.2732293}
{doi: {{%
10\hspace{.1pt}\discretionary{.}{%
}{.}\hspace{.4pt}14778\discretionary{/}{%
}{/}2732286\hspace{.1pt}\discretionary{.}{%
}{.}\hspace{.4pt}2732293}}}


\bibitem{ye2024gnnpe}
Y.~Ye, X.~Lian, and M.~Chen.
\newblock Efficient exact subgraph matching via gnn-based path dominance embedding.
\newblock {\em Proceedings of the VLDB Endowment}, 17(7):1628--1641, Mar. 2024. \href{https://doi.org/10.14778/3654621.3654630}
{doi: {{%
10\hspace{.1pt}\discretionary{.}{%
}{.}\hspace{.4pt}14778\discretionary{/}{%
}{/}3654621\hspace{.1pt}\discretionary{.}{%
}{.}\hspace{.4pt}3654630}}}


\bibitem{yi2017autog}
P.~Yi, B.~Choi, S.~S. Bhowmick, and J.~Xu.
\newblock Autog: A visual query autocompletion framework for graph databases.
\newblock {\em The VLDB Journal}, 26(3):347--372, Jan. 2017. \href{https://doi.org/10.1007/s00778-017-0454-9}
{doi: {{%
10\hspace{.1pt}\discretionary{.}{%
}{.}\hspace{.4pt}1007\discretionary{/}{%
}{/}s00778\discretionary{%
}{-}{-}017\discretionary{%
}{-}{-}0454\discretionary{%
}{-}{-}9}}}


\bibitem{zhang2009gaddi}
S.~Zhang, S.~Li, and J.~Yang.
\newblock {GADDI}: Distance index based subgraph matching in biological networks.
\newblock In {\em Proceedings of the 12th International Conference on Extending Database Technology: Advances in Database Technology}, pp. 192--203. ACM, Mar. 2009. \href{https://doi.org/10.1145/1516360.1516384}
{doi: {{%
10\hspace{.1pt}\discretionary{.}{%
}{.}\hspace{.4pt}1145\discretionary{/}{%
}{/}1516360\hspace{.1pt}\discretionary{.}{%
}{.}\hspace{.4pt}1516384}}}


\bibitem{zhao2024malicious}
Y.~Zhao, S.~Lv, W.~Long, Y.~Fan, J.~Yuan, H.~Jiang, and F.~Zhou.
\newblock Malicious webshell family dataset for webshell multi-classification research.
\newblock {\em Visual Informatics}, 8(1):47--55, Mar. 2024. \href{https://doi.org/10.1016/j.visinf.2023.06.008}
{doi: {{%
10\hspace{.1pt}\discretionary{.}{%
}{.}\hspace{.4pt}1016\discretionary{/}{%
}{/}j\hspace{.1pt}\discretionary{.}{%
}{.}\hspace{.4pt}visinf\hspace{.1pt}\discretionary{.}{%
}{.}\hspace{.4pt}2023\hspace{.1pt}\discretionary{.}{%
}{.}\hspace{.4pt}06\hspace{.1pt}\discretionary{.}{%
}{.}\hspace{.4pt}008}}}


\bibitem{zhou24adamotif}
H.~Zhou, P.~Lai, Z.~Sun, X.~Chen, Y.~Chen, H.~Wu, and Y.~Wang.
\newblock {AdaMotif}: Graph simplification via adaptive motif design.
\newblock {\em IEEE Transactions on Visualization and Computer Graphics}, 31(1):688–698,  11 pages, Sept. 2024. \href{https://doi.org/10.1109/TVCG.2024.3456321}
{doi: {{%
10\hspace{.1pt}\discretionary{.}{%
}{.}\hspace{.4pt}1109\discretionary{/}{%
}{/}TVCG\hspace{.1pt}\discretionary{.}{%
}{.}\hspace{.4pt}2024\hspace{.1pt}\discretionary{.}{%
}{.}\hspace{.4pt}3456321}}}


\end{thebibliography}

\end{document}